\documentclass[12pt,preprint]{aastex}

\newcommand{\HESS}{H.E.S.S.\,}
\newcommand{\gr}{$\gamma$-ray}
\newcommand{\grs}{$\gamma$-rays}
\newcommand{\vhe}{V\textsc{HE}}
\newcommand{\dg}{\ensuremath{^\circ}}

\slugcomment{Accepted by the Astrophysical Journal}

\shorttitle{The H.E.S.S. survey of the inner Galaxy}
\shortauthors{F.~A. Aharonian et al.}

\begin{document}

\title{The \HESS\  survey of the Inner Galaxy in very-high-energy
gamma-rays}



\author{F. Aharonian\altaffilmark{1},
 A.G.~Akhperjanian \altaffilmark{2},
 A.R.~Bazer-Bachi \altaffilmark{3},
 M.~Beilicke \altaffilmark{4},
 W.~Benbow \altaffilmark{1},
 D.~Berge \altaffilmark{1},
 K.~Bernl\"ohr \altaffilmark{1,5},
 C.~Boisson \altaffilmark{6},
 O.~Bolz \altaffilmark{1},
 V.~Borrel \altaffilmark{3},
 I.~Braun \altaffilmark{1},
 F.~Breitling \altaffilmark{5},
 A.M.~Brown \altaffilmark{7},
 P.M.~Chadwick \altaffilmark{7},
 L.-M.~Chounet \altaffilmark{8},
 R.~Cornils \altaffilmark{4},
 L.~Costamante \altaffilmark{1,20},
 B.~Degrange \altaffilmark{8},
 H.J.~Dickinson \altaffilmark{7},
 A.~Djannati-Ata\"i \altaffilmark{9},
 L.O'C.~Drury \altaffilmark{10},
 G.~Dubus \altaffilmark{8},
 D.~Emmanoulopoulos \altaffilmark{11},
 P.~Espigat \altaffilmark{9},
 F.~Feinstein \altaffilmark{12},
 G.~Fontaine \altaffilmark{8},
 Y.~Fuchs \altaffilmark{13},
 S.~Funk \altaffilmark{1, *},
 Y.A.~Gallant \altaffilmark{12},
 B.~Giebels \altaffilmark{8},
 S.~Gillessen \altaffilmark{1},
 J.F.~Glicenstein \altaffilmark{14},
 P.~Goret \altaffilmark{14},
 C.~Hadjichristidis \altaffilmark{7},
 M.~Hauser \altaffilmark{11},
 G.~Heinzelmann \altaffilmark{4},
 G.~Henri \altaffilmark{13},
 G.~Hermann \altaffilmark{1},
 J.A.~Hinton \altaffilmark{1},
 W.~Hofmann \altaffilmark{1},
 M.~Holleran \altaffilmark{15},
 D.~Horns \altaffilmark{1},
 A.~Jacholkowska \altaffilmark{12},
 O.C.~de~Jager \altaffilmark{15},
 B.~Kh\'elifi \altaffilmark{1},
 Nu.~Komin \altaffilmark{5},
 A.~Konopelko \altaffilmark{1,5},
 I.J.~Latham \altaffilmark{7},
 R.~Le Gallou \altaffilmark{7},
 A.~Lemi\`ere \altaffilmark{9},
 M.~Lemoine-Goumard \altaffilmark{8},
 N.~Leroy \altaffilmark{8},
 T.~Lohse \altaffilmark{5},
 J.M.~Martin \altaffilmark{6},
 O.~Martineau-Huynh \altaffilmark{16},
 A.~Marcowith \altaffilmark{3},
 C.~Masterson \altaffilmark{1,20},
 T.J.L.~McComb \altaffilmark{7},
 M.~de~Naurois \altaffilmark{16},
 S.J.~Nolan \altaffilmark{7},
 A.~Noutsos \altaffilmark{7},
 K.J.~Orford \altaffilmark{7},
 J.L.~Osborne \altaffilmark{7},
 M.~Ouchrif \altaffilmark{16,20},
 M.~Panter \altaffilmark{1},
 G.~Pelletier \altaffilmark{13},
 S.~Pita \altaffilmark{9},
 G.~P\"uhlhofer \altaffilmark{1,11},
 M.~Punch \altaffilmark{9},
 B.C.~Raubenheimer \altaffilmark{15},
 M.~Raue \altaffilmark{4},
 J.~Raux \altaffilmark{16},
 S.M.~Rayner \altaffilmark{7},
 A.~Reimer \altaffilmark{17},
 O.~Reimer \altaffilmark{17},
 J.~Ripken \altaffilmark{4},
 L.~Rob \altaffilmark{18},
 L.~Rolland \altaffilmark{16},
 G.~Rowell \altaffilmark{1},
 V.~Sahakian \altaffilmark{2},
 L.~Saug\'e \altaffilmark{13},
 S.~Schlenker \altaffilmark{5},
 R.~Schlickeiser \altaffilmark{17},
 C.~Schuster \altaffilmark{17},
 U.~Schwanke \altaffilmark{5},
 M.~Siewert \altaffilmark{17},
 H.~Sol \altaffilmark{6},
 D.~Spangler \altaffilmark{7},
 R.~Steenkamp \altaffilmark{19},
 C.~Stegmann \altaffilmark{5},
 J.-P.~Tavernet \altaffilmark{16},
 R.~Terrier \altaffilmark{9},
 C.G.~Th\'eoret \altaffilmark{9},
 M.~Tluczykont \altaffilmark{8,20},
 G.~Vasileiadis \altaffilmark{12},
 C.~Venter \altaffilmark{15},
 P.~Vincent \altaffilmark{16},
 H.J.~V\"olk \altaffilmark{1},
 S.J.~Wagner \altaffilmark{11},
}

\altaffiltext{*}{Correspondence and request for material should be
sent to Stefan.Funk@mpi-hd.mpg.de} \altaffiltext{1}{
Max-Planck-Institut f\"ur Kernphysik, Heidelberg, Germany}
\altaffiltext{2}{Yerevan Physics Institute, Yerevan, Armenia}
\altaffiltext{3}{Centre d'Etude Spatiale des Rayonnements, CNRS/UPS,
Toulouse, France} \altaffiltext{4}{Universit\"at Hamburg, Institut
f\"ur Experimentalphysik, Hamburg, Germany} \altaffiltext{5}{Institut
f\"ur Physik, Humboldt-Universit\"at zu Berlin, Germany}
\altaffiltext{6}{LUTH, UMR 8102 du CNRS, Observatoire de Paris,
Section de Meudon, France} \altaffiltext{7}{University of Durham,
Department of Physics, Durham DH1 3LE, U.K.}
\altaffiltext{8}{Laboratoire Leprince-Ringuet, IN2P3/CNRS, Ecole
Polytechnique, Palaiseau, France} \altaffiltext{9}{APC, Paris Cedex
05, France (7164 (CNRS, Observatoire de Paris))}
\altaffiltext{10}{Dublin Institute for Advanced Studies, Dublin,
Ireland} \altaffiltext{11}{Landessternwarte, K\"onigstuhl, Heidelberg,
Germany} \altaffiltext{12}{Laboratoire de Physique Th\'eorique et
Astroparticules, IN2P3/CNRS, Universit\'e Montpellier II, France} 
 \altaffiltext{13}{Laboratoire d'Astrophysique de Grenoble,
INSU/CNRS, Universit\'e Joseph Fourier, France}
\altaffiltext{14}{DAPNIA/DSM/CEA, CE Saclay, Gif-sur-Yvette, France}
\altaffiltext{15}{Unit for Space Physics, North-West University,
Potchefstroom, South Africa} \altaffiltext{16}{Laboratoire de Physique
Nucl\'eaire et de Hautes Energies, Universit\'es Paris VI \& VII,
France} \altaffiltext{17}{Institut f\"ur Theoretische Physik,
Lehrstuhl IV, Ruhr-Universit\"at Bochum, Germany}
\altaffiltext{18}{Institute of Particle and Nuclear Physics, Charles
University, Prague, Czech Republic} \altaffiltext{19}{University of
Namibia, Windhoek, Namibia} \altaffiltext{20}{European Associated
Laboratory for Gamma-Ray Astronomy, jointly supported by CNRS and MPG}

\begin{abstract}
We report on a survey of the inner part of the Galactic Plane in very
high energy \grs\, with the \HESS\ Cherenkov telescope system. The
Galactic Plane between $\pm 30$\dg\ in longitude and $\pm 3$\dg\ in
latitude relative to the Galactic Centre was observed in 500 pointings
for a total of 230~hours, reaching an average flux sensitivity of 2\%
of the Crab Nebula at energies above 200~GeV. Fourteen previously
unknown sources were detected at a significance level greater than
4$\sigma$ after accounting for all trials involved in the
search. Initial results on the eight most significant of these sources
were already reported elsewhere~\citep{HESSScan}. Here we present
detailed spectral and morphological information for all the new sources,
along with a discussion on possible counterparts in other wavelength
bands. The distribution in Galactic latitude of the detected sources
appears to be consistent with a scale height in the Galactic disk for
the parent population smaller than 100~pc, consistent with
expectations for supernova remnants and/or pulsar wind nebulae.
\end{abstract}


\section{Introduction}
Very high energy (\vhe, defined here as $E >\,10^{11}$~eV) \grs\,
provide the most direct view currently available into extreme
environments in the local universe.  As \grs\, are produced by
interactions of relativistic particles they represent a good probe of
non-thermal astrophysical processes. Relativistic electrons can
produce \grs\, by non-thermal bremsstrahlung or by inverse Compton
scattering on background radiation fields, whereas protons (and atomic
nuclei) can generate \grs\, via the decay of $\pi^{0}$s produced in
hadronic interactions with ambient material. \vhe\, $\gamma$-radiation
is therefore expected from the acceleration sites of particles with
energies above 1~TeV ($10^{12}$~eV). In particular, \grs\, can be seen
as a probe of sources of nucleonic cosmic rays in our Galaxy and may
thus provide part of the solution to the century old puzzle of the
origin of this radiation.

Observations of synchrotron emission in the X-ray band provide strong
indirect evidence for the existence of $\sim100$~TeV electrons in
shell-type supernova remnants (SNR), see for
example~\citet{SN1006Koyama},
and in pulsar wind nebulae (PWN), see e.g. ~\citet{ChandraCrab}.
At TeV energies, shell-type SNRs and pulsars are established to be
sources of \vhe\, \grs\, as shown by highly significant detections of
the Crab Nebula by the Whipple collaboration~\citep{CrabWhipple} or
the detection of the Supernova remnant RX\,J1713.7--3946 by the
CANGAROO~\citep{RXJCangaroo} and the \HESS\
collaborations~\citep{HESSRXJ}. However, despite the fact that SNRs
are the best candidates for the acceleration of hadronic cosmic rays
up to the so-called \emph{knee} in the cosmic ray particle spectrum at
$\sim10^{15}$~eV, there is as yet no generally accepted evidence for
the existence of energetic nuclei (as opposed to electrons) in these
objects. SNRs are favoured as the site of cosmic-ray acceleration for
two principal reasons; first, that a theoretically well established
acceleration mechanism exists in the form of diffusive shock
acceleration at SNR shock fronts, for reviews see
e.g.~\citet{Blandford, Malkov}, and secondly, that the Galactic
supernovae and the resulting SNRs are the only known potential sources
which can provide the necessary amount of energy~\citep{Ginzburg}.

To explain the observed energy density of cosmic rays, supernova
explosions must convert kinetic energy to energy of relativistic
particles with a conversion efficiency $\Theta$ of roughly $0.1 -
0.2$. Theoretical expectations for the \gr\,-emission of SNRs arising
from hadronic interactions are given by~\citet{DrurySNR} as
\begin{equation}
  \label{equation::SNR}
  F_{\gamma}(> E) \sim 9 \times 10^{-11}\,\Theta\,\left(
      \frac{E}{\mbox{1 TeV}}\right)^{-1.1}
      \times\,S\,\mbox{cm}^{-2}\mbox{s}^{-1}
\end{equation}
with
\begin{equation}
 S =\left(\frac{E_{\mathrm{SN}}}{10^{51}\, \mbox{erg}}\right) \left(\frac{d}{1
  \, \mbox{kpc}}\right)^{-2} \left(\frac{n}{1\, \mbox{cm}^{-3}}\right)
\end{equation}
where $d$ denotes the distance to the object, $n$ the density of the
circumstellar medium assumed to be uniform, and $E_{\mathrm{SN}}$ the
total kinetic energy released by the Supernova. This formula assumes a
power-law distribution of cosmic-ray energies with a typical
differential photon index of 2.1. Calculations of the photon index
give a range of values around the index value 2.0 for a uniform
circumstellar medium~\citep{BerezhkoVoelk1997} and a rather more
extended range of photon index values for Supernova explosions into
the wind bubble of a very massive progenitor
star~\citep{BerezhkoVoelk2000}.

A typical value for the fraction of kinetic energy converted into
cosmic rays is $\Theta\,E_{\mathrm{SN}} =
1.0\times10^{50}$~erg~\citep{Malkov, Voelk}. For an SNR at a distance
of $d = 1$~kpc expanding into an environment with a typical density of
the interstellar medium of $n = 1$ hydrogen atom per cm$^3$, a \vhe\,
\gr\, flux of 20\% of that from the Crab Nebula is expected. The
luminosity $L_0$ of such a source is $7.2\times 10^{33}$~erg~s$^{-1}$ in
the energy range between 0.2~TeV and 10~TeV. One would therefore
expect that with the sensitivity level of the \HESS\ Galactic Plane
survey (down to $\sim2$\% of the Crab flux) SNRs should be detectable
out to a distance of 3~kpc (to be compared to 8.5~kpc to the centre of
the Galaxy). For larger values of $n$, SNRs should be visible within
most of the volume of the Galactic Disc.

Another potential source of \vhe\, \grs\, are pulsars and
PWN. Pulsars, rapidly rotating neutron stars left over after a
supernova explosion, are highly magnetised and act like unipolar
inductors accelerating particles. The pulsar wind interacts with the
ambient medium, generating a shock region where particles are
accelerated. Such objects may therefore exhibit a pulsed component of
radiation, from the immediate vicinity of the pulsar, together with an
unpulsed component from the shock region and beyond. The Crab Nebula
is the best-studied object of this class of particle accelerators.

Most of these potential Galactic \gr\, sources such as SNRs and
pulsars are associated with the formation of massive stars and
therefore cluster along the Galactic Plane, especially concentrated in
the direction towards the centre of our Galaxy. 91 SNRs and 389
pulsars are catalogued within the inner 60\dg\ in Galactic longitude
and 6\dg\ in Galactic latitude~\citep{Green,ATNF}. Despite the large
number of known SNRs and pulsars in our Galaxy, the actual number of
detected \vhe\, \gr\, sources was until now limited to only a handful
of objects (for a review, see e.g.~\citet{Weekes}).

Particle acceleration to multi-TeV energies may also be powered by OB
stellar associations, via the {\emph{superbubble}} mechanism
(see~\citet{Parizot} for a summary), and also, possibly more directly
via the stellar winds of member stars and subsequent interaction (see
e.g.~\citet{Montmerle, Voelk1982, White1985, Benaglia2001}). There are
presently two TeV sources that may be linked to such OB associations.
TeV\,J2032+4130, discovered by the HEGRA
collaboration~\citep{HEGRAUnid}, and HESS\,J1303--631, a recent
discovery by H.E.S.S.~\citep{HESS1303} are in the vicinity of the OB
associations Cygnus~OB2 and Centaurus~OB1, respectively.  Both TeV
sources remain unidentified (UID) in the sense that there are as yet
no clear counterparts of similar size and morphology in other
wavebands. These sources might represent a new class of objects
distributed along the Galactic Plane.

At lower energies, observations by the space based instrument EGRET in
the GeV energy range~\citep{EGRET} also revealed a population of
sources clustering along the Galactic Plane. The limited angular
resolution of the EGRET instrument renders an identification of these
sources in other wavebands very difficult and most of these objects
remain as yet unidentified.

For all these reasons, a survey of the inner part of the Galaxy
provides an efficient way to search for \gr\, emission in order to
investigate properties of source classes and search for yet unknown
types of Galactic \vhe\, \gr\, emitters.

At TeV energies the Milagro water-Cherenkov detector
~\citep{MilagroAllSky} and the Tibet air-shower
array~\citep{TibetAllSky} have been used to perform large-scale
surveys. While these instruments have the advantage of a very wide
field of view ($\sim1$~sterad), the sensitivity obtained by these
surveys is rather limited, reaching a flux limit comparable to the
flux of the Crab Nebula, $\sim3 \times 10^{-10}$~cm$^{-2}$~s$^{-1}$
(for $E >200\,$~GeV), in one year of observations. Both surveys
covered $\sim2 \pi$~sterad of the northern sky and revealed no
previously unknown significant \gr\, sources.  However, some evidence
for \gr\, emission was found by the Milagro Collaboration from a
region in the Cygnus constellation and another region close to the
Crab Nebula~\citep{Milagro_HS}. Recently the detection of diffuse
emission from the Galactic Plane with Milagro has been
reported~\citep{Milagro_Diffuse}.

The HEGRA instrument was the first array of imaging air Cherenkov
telescopes to be used to survey a part of the Galactic
Plane~\citep{HEGRASCAN}. The range of Galactic longitudes ($l$)
$-2^{\circ} < l < 85^{\circ}$ was observed; due to the location of
HEGRA in the northern hemisphere and the resulting large zenith angles
for observations of the centre of the Galactic Plane, the sensitivity
was reduced for the central part of the Galaxy and the energy
threshold of the observations ranged from 500~GeV to 7~TeV. No
sources of \vhe\, \grs\, were found in this survey, and upper limits
between 15\% of the Crab flux for Galactic longitudes $l>30^{\circ}$
and more than 30\% of the Crab flux in the inner part of the Milky Way
were derived.

Until the completion of the High Energy Stereoscopic System
(\HESS)~\citep{HESS} in early 2004, no \vhe\, \gr\, survey of
comparable sensitivity of the southern sky, or of the central region
of the Galaxy has been performed. In this paper we describe a survey
of this region with \HESS\ between May and July 2004 at a flux
sensitivity down to 2\% of the Crab flux. The initial results from the
survey including the description of the discovery of eight new sources
have been summarised in~\citet{HESSScan}. This work presents detailed
results on these eight sources. In addition we report the detection of
six additional weaker sources of \vhe\, \grs\, extending to lower flux
levels than our previous analysis. Spectral information and a more
detailed study on the morphologies for all sources is provided. The
paper is structured as follows: after a description of the
H.E.S.S. instrument (section~\ref{instrument}) and the Galactic survey
dataset used in this paper (section~\ref{dataset}), the data analysis
technique is presented in
section~\ref{analysis}. Section~\ref{characteristics} gives a detailed
description of the individual sources along with possible counterparts
in other wavebands, followed by a discussion of the results in
section~\ref{discussion} and some conclusions in
section~\ref{conclusion}.

\section{The \HESS\  Instrument}
\label{instrument}
\HESS\ is a system of four imaging atmospheric Cherenkov telescopes,
located in the Khomas Highlands of Namibia at a height of 1800~m above
sea level. \HESS\ has unprecedented sensitivity in the energy range
from 100~GeV to several tens of TeV. The imaging Cherenkov technique
exploits the cascade of secondary particles produced in interactions
of primary \grs\, with molecules in the Earth's upper
atmosphere. These secondary particles emit Cherenkov light, which can
be imaged by the Cherenkov telescopes. The energy and direction of the
primary \grs\, can be reconstructed based on the intensity and shape
of the detected Cherenkov light distribution.

Multiple views of the same particle shower with several telescopes
improve background rejection capability and the angular and energy
resolution of the instrument, in comparison to a single
telescope. Therefore, \HESS\ observations are performed in a
stereoscopic mode, whereby at least two of the four telescopes must be
triggered by a shower for an event to be recorded~\citep{HESSTrigger}.
The telescopes each have a mirror area of
107~m$^{2}$~\citep{HESSOptics, HESSOpticsII}. The Cherenkov light is
focussed onto fast 960~pixel photo-multiplier cameras, which record
the Cherenkov light images~\citep{HESSCamera}. The 5\dg\ diameter
field of view of the \HESS\ cameras is the largest of all imaging
Cherenkov detectors and is especially suited for the study of extended
sources and the search for unknown sources in surveys. This was part
of the original design concept for the \HESS\ telescope
system~\citep{Aharonian1997a, Aharonian1997b}.  

The \HESS\ instrument has an energy threshold of $\sim100$~GeV at
zenith before selection cuts, an angular resolution of
$\sim0.1^{\circ}$ for individual $\gamma$-rays and a point source
sensitivity of $<2.0\times\,10^{-13}$~cm$^{-2}$~s$^{-1}$ above 1~TeV
(1\% of the flux from the Crab Nebula) for a 5$\sigma$ detection in a
25~hour observation. The sensitivity of \HESS\ and the capability to
map large extended sources has been demonstrated by the resolved
images of the SNRs RX\,J1713.7--3946~\citep{HESSRXJ} and
RX\,J0852.0--4622~\citep{HESSVela}. The serendipitous detections of
the unidentified source HESS\,J1303--631 in the field of view of the
binary pulsar PSR\,B1259--63 and of the composite supernova remnant
G\,0.9+0.1~\citep{HESSG0.9} in the field of view of the Galactic
Centre source HESS\,J1745--290~\citep{HESSGalCen} demonstrate the
advantage of a large field of view. This large field of view implies
that some sources are observed up to 2 degrees off the optical axis of
the telescope system.

\begin{figure}
  \epsscale{.90}           
  \plotone{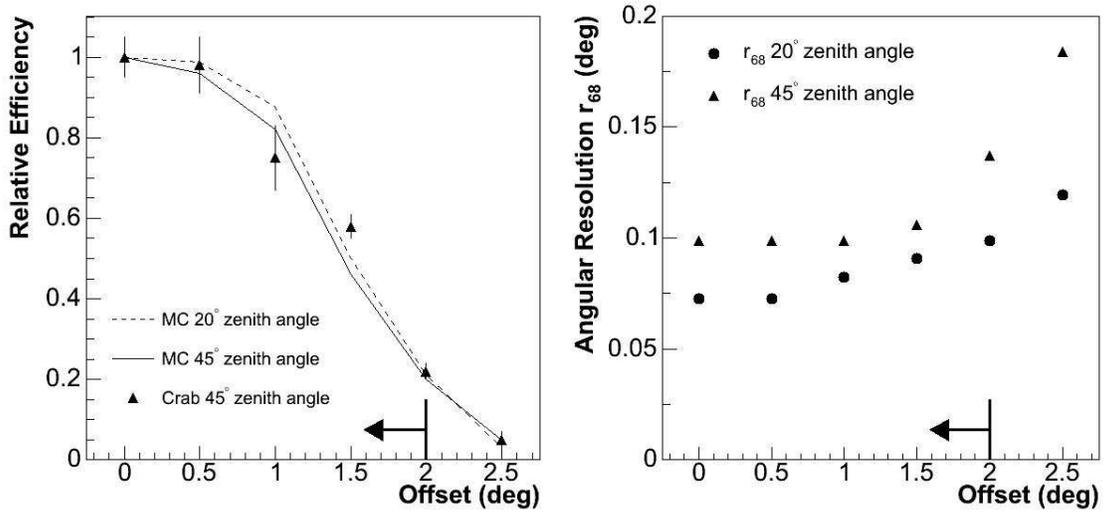}
  \caption{The off-axis performance of the \HESS\ instrument.  The
    left figure shows the \gr\, efficiency relative to that for an
    on-axis source for Monte-Carlo simulations (dashed line for 20\dg\
    zenith angle and solid line for 45\dg\ zenith angle) and for
    H.E.S.S. observations of the Crab Nebula at a mean zenith angle of
    47\dg\ (triangles). The Monte-Carlo simulations are generated
    assuming a Crab-like source spectrum of the form
    $dN/dE\,\propto\,E^{-\Gamma}$ with $\Gamma = 2.5$.  The right
    figure shows the simulated angular resolution (68\% containment
    radius) as a function of the offset for observations at zenith
    angles of 20\dg\ (circles) and 50\dg\ (triangles),
    respectively. In each plot the arrow denotes the offset of $2\dg$
    up to which \grs\, are included in this analysis.
    \label{fig::Offset}}          
\end{figure}

To illustrate the off-axis performance within the field of view of
H.E.S.S., Figure~\ref{fig::Offset} summarises some basic parameters as
a function of the offset. The left panel shows the \gr\, efficiency
relative to on-axis pointing, derived from Monte-Carlo simulations for
two different zenith angles along with measurements of the Crab
Nebula. The plot shows a weak dependence of the \gr\, efficiency on
zenith angle and reasonable agreement between Monte-Carlo simulations
and data. The right panel of Figure~\ref{fig::Offset} shows the
angular resolution of \HESS\ (specified by the 68\% containment radius
of the point spread function (PSF)) versus offset in the camera for
observations at two different zenith angles. It is evident that the
angular resolution is better than 0.1\dg\ and is reasonably constant
in the inner 1.5\dg\ of the field of view.
\section{The \HESS\  Galactic Plane survey}
\label{dataset}
\begin{figure}
  \epsscale{.9}           
  \plotone{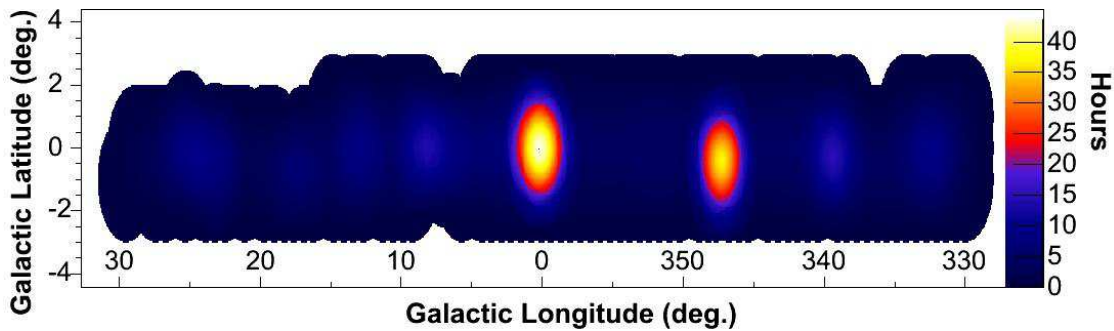}
  \caption{The effective exposure of the \HESS\  Galactic Plane
    survey, expressed as equivalent number of hours spent observing a
    source at the centre of the field of view. The deeper observations
    of the RX\,J1713.7--3946 and the Galactic Centre are apparent.
    \label{fig::AcceptanceTime}}          
\end{figure}

\begin{figure}
  \epsscale{.6}           
  \plotone{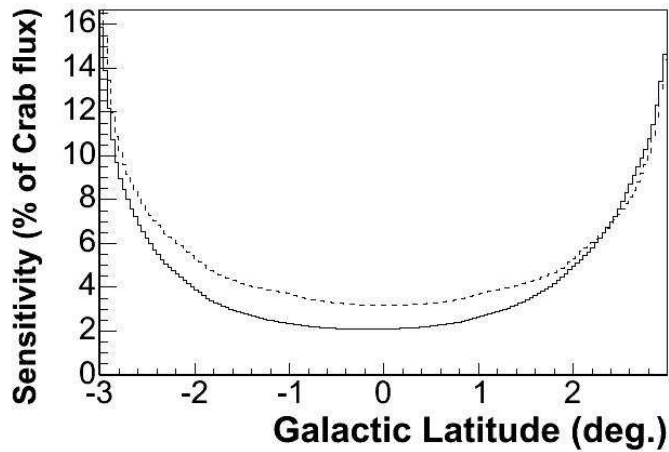}
  \caption{Sensitivity in Galactic latitude of the \HESS\ Galactic
    Plane scan in units of the flux from the Crab Nebula ($\sim3
    \times 10^{-10}$~cm$^{-2}$~s$^{-1}$, $E >200\,$~GeV) and assuming
    a power-law spectrum with photon index $\Gamma = 2.5$. The
    sensitivity used here is defined as the minimum flux of a
    point-source that would be detectable in this survey by \HESS\ at
    the 6.3$\sigma$ level (corresponding to 4$\sigma$ after accounting
    for all trials), given the cuts used to detect slightly extended
    sources (in particular a $\theta^2 = 0.05$ deg$^2$ cut). The solid
    curve shows the sensitivity averaged over the whole longitude
    range of the survey, including runs on the Galactic Centre region
    and on the SNR RX\,J1713.7--3946. The dashed curve shows the
    sensitivity in the range of longitude $350\dg\ < l < 357\dg$, a
    region where no re-observations were conducted.
  \label{fig::Sensitivity}}          
\end{figure}

The survey was conducted in the Galactic longitude band $\pm
30^{\circ}$ around $l = 0^{\circ}$. Runs of 28~minutes duration were
taken at pointing positions with a spacing of 0.7\dg\ in longitude in
three strips in Galactic latitude: $b = -1^{\circ}, b = 0^{\circ}$ and
$b = +1^{\circ}$. For the data analysis runs were selected based on
weather and hardware conditions. 95~hours of quality-selected data
were taken in the scan mode between May and July 2004. The observation
schedule was optimised so that individual pointings were conducted
near culmination, resulting in the smallest zenith angles
accessible. Promising source candidates were re-observed between July
and October 2004, yielding a further 30 good-quality hours of
data. For the analysis described in the following, additional pointed
observations of SNR RX\,J1713.7--3946 and of the Galactic Centre
region are included. The total data set presented here amounts to
230~hours after quality selection, the mean zenith angle of all
observations is 26\dg.  Figure \ref{fig::AcceptanceTime} shows a map
of exposure for this data set, expressed in terms of an equivalent
number of hours observing a source at the centre of the field of view.
As can be seen from Figure \ref{fig::AcceptanceTime}, the exposure is
not uniform with peak exposures in the regions around
RX\,J1713.7--3946 and the Galactic Centre. This irregular exposure
results in an uneven sensitivity for the detection of new sources. The
sensitivity of the survey in Galactic latitude (averaged over the
whole longitude range) is shown in Figure~\ref{fig::Sensitivity}.  For
$-1.5\dg\ < b < 1.5\dg$ the sensitivity is rather flat at a level of
2\% of the flux from the Crab Nebula, and deteriorates rapidly at
$>2$\dg\ from the Galactic Plane. The flux sensitivity above 200 GeV
is $\sim50$\% worse for a source with photon index $\Gamma = 3.0$
compared to one with $\Gamma = 2.0$.

\section{Data Analysis}
\label{analysis}
Following calibration of the data~\citep{HESSCalib}, the standard
\HESS\ event reconstruction scheme was applied to the dataset. After
tail-cut image cleaning and a Hillas parametrisation algorithm, the
direction of the shower axis was reconstructed by intersection of the
major axes of the Hillas ellipses. \HESS\ standard
cuts~\citep{HESS2155} on the width and length of these ellipses
(scaled to the width and length expected from Monte-Carlo simulations)
were used to suppress the hadronic background. An additional cut on
the image size $> 200$~photo-electrons (p.e.) yields optimum
sensitivity for weak sources provided that source spectra are not
significantly steeper than that of the Crab Nebula. They result in an
angular resolution of 0.08\dg\ (68\% containment radius) and an energy
threshold of 250~GeV.
To reduce systematic effects which affect images close to the edge of
the camera, only events reconstructed within a maximum distance of
2$^{\circ}$ from the pointing direction of the system were used for
this analysis. The survey region was divided (a priori) into a grid of
trial source positions with grid points spaced by 0.04\dg, a value
well below the angular resolution of the instrument.
For each grid point in the scan region, the on-source signal was
calculated by summing up all events within a circle of radius
$\theta=0.22^{\circ}$, (or $\theta^2=0.05$ deg$^2$), where $\theta$
denotes the angular distance between the reconstructed event direction
and the grid point. This cut is appropriate for slightly extended
sources with respect to the angular resolution of the instrument and
was chosen a-priori, since it is optimal for the detection of \gr\,
sources with the average observed size of Galactic
SNRs~\citep{Green}. A search for very extended sources (relative to
the angular resolution) using a larger $\theta^2$ cut yields similar
results. A similar search for point-like sources with a $\theta^2 =
0.01$ deg$^2$ cut revealed a subset of the sources presented here and,
additionally, the point-like source HESS\,J1826--148, which is
associated with the microquasar LS5039. The discovery of this
source is reported elsewhere~\citep{HESSLS5039}.

\begin{figure}
  \epsscale{.65}           
  \plotone{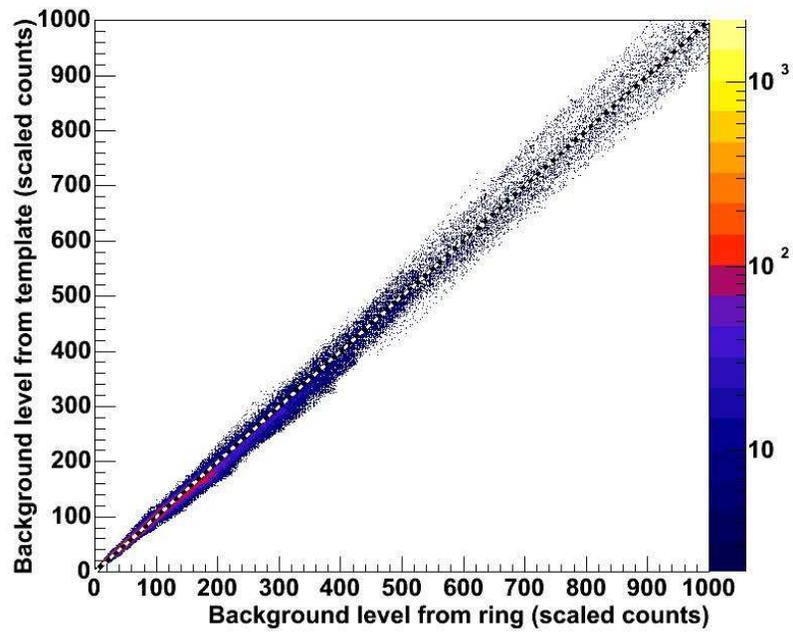}
  \caption{Background level estimated for each grid point, calculated
  using a ring background technique, plotted against the background
  level estimated from a template method~\citep{Template}. The black
  dashed line shows the expected one-to-one correlation.
  \label{fig::Background}}          
\end{figure}

\begin{figure}
  \epsscale{.80}           
  \plotone{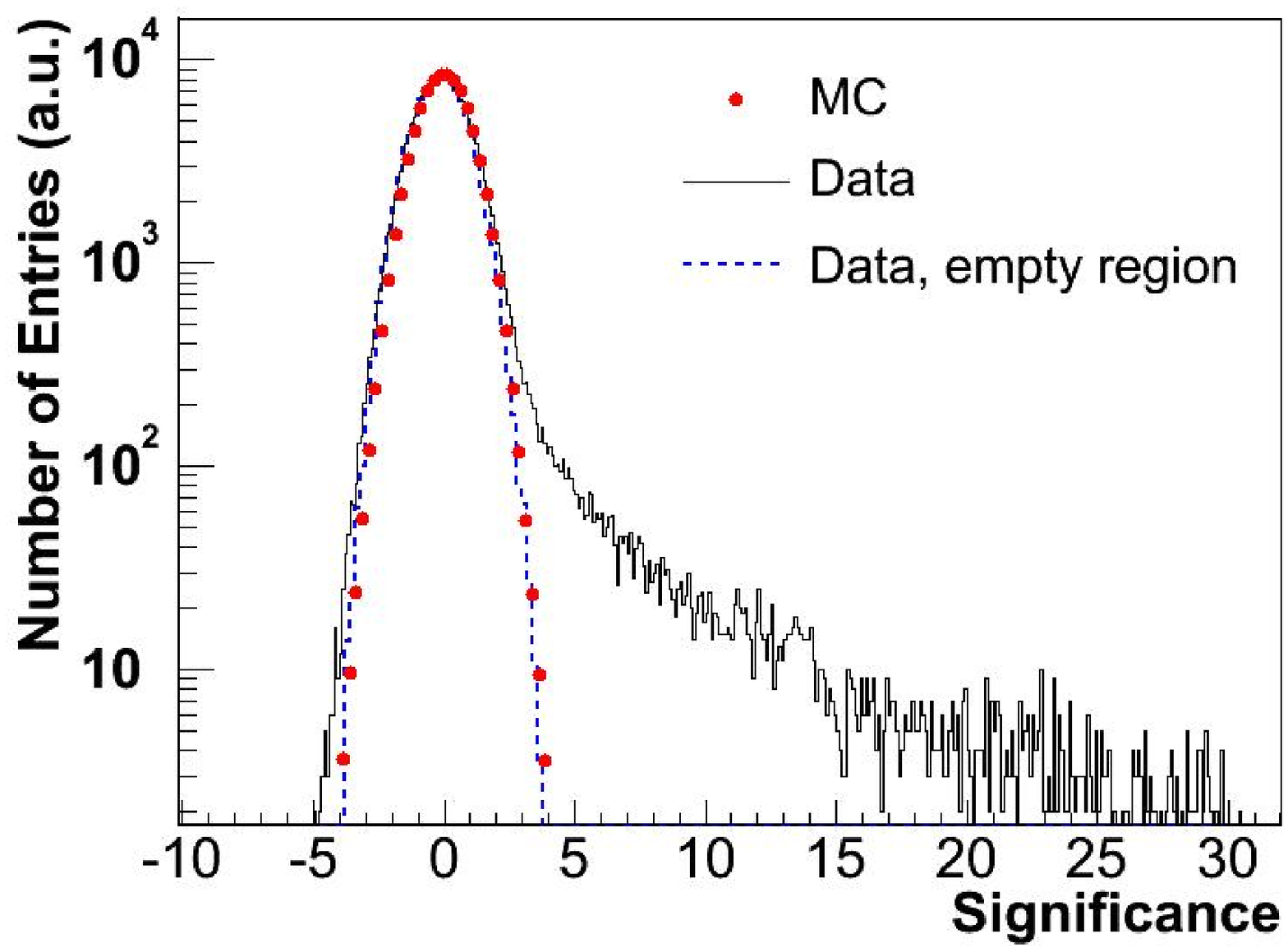}
  \caption{Distribution of significances in the survey region (solid
    black). For comparison, the distribution of significances in a
    region without \gr\, sources is shown in dashed blue. The
    distribution of significances for a Monte-Carlo simulation of the
    survey without $\gamma$-ray sources is shown as red points and
    follows a Gaussian distribution with zero mean and unit width. A
    Gaussian fit to the dashed blue distribution yields a mean of
    $-0.050 \pm 0.003$ with an RMS of $1.027 \pm 0.001$, indicating a
    small systematic error in the determination of the significances
    at the level of a few percent. The dashed blue curve and the red
    data points have been scaled to the peak height of the black
    distribution.
  \label{fig::SignificanceDistribution}}          
\end{figure}

\begin{figure}
  \plotone{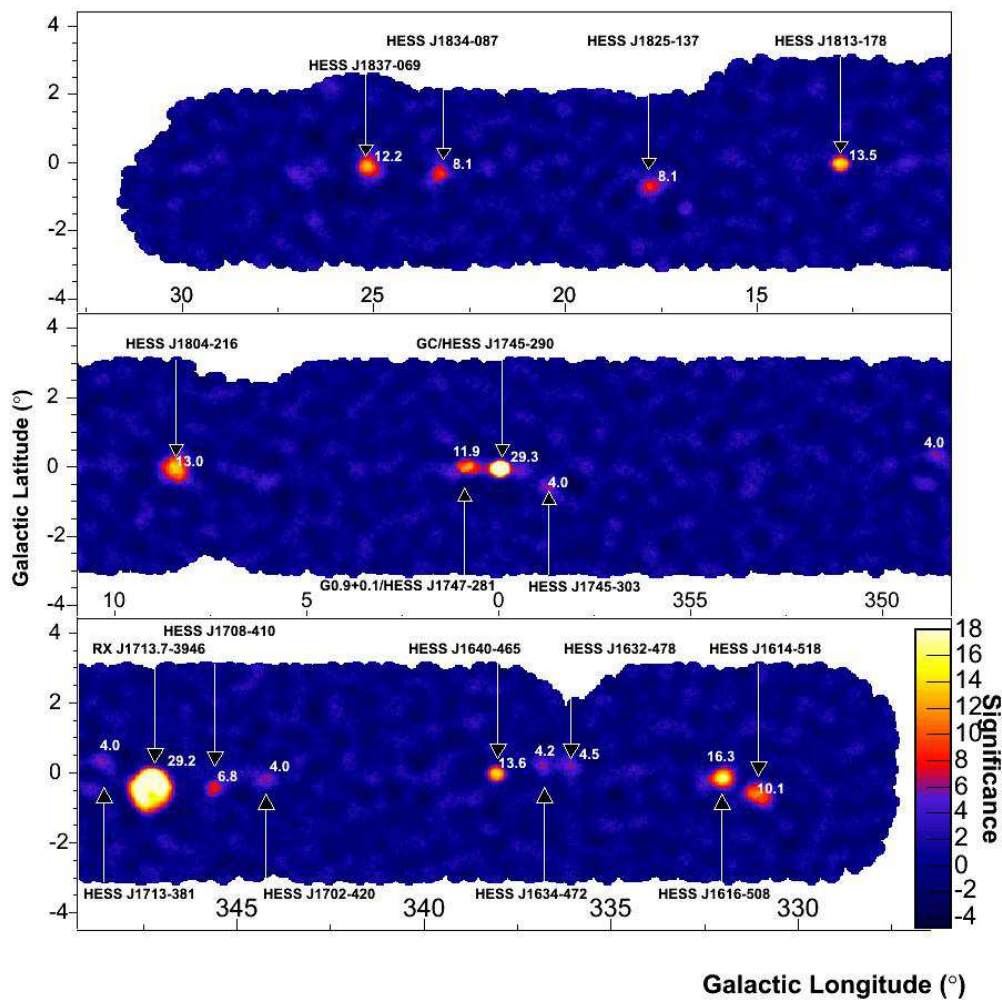}
  \caption{Significance Map of the \HESS\ Galactic Plane survey in
  2004, including data from re-observations of source candidates
  detected in the original scan and observations of the known \gr\,
  emitter RX\,J1713.7--3946 and the Galactic Centre region. The
  typical energy threshold for this map is 250~GeV. The on-source
  counts for each grid point are integrated in a circle of radius
  $\theta = 0.22^{\circ}$. The background for each grid point has been
  derived using a ring of mean radius 0.6\dg\ and an area 7 times
  that of the on-source circle. The labels indicate the \gr\, sources
  described in this work, along with the known \gr\, sources
  RX\,J1713.7--3946 (HESS\,J1713--397), G\,0.9+0.1 (HESS\,J1747--281)
  and the Galactic Centre (HESS\,J1745--290). The numbers in the map
  give the post-trials significances of the \gr\, sources. The
  significance scale is truncated at 18$\sigma$, the signals from the
  Galactic Centre and RX\,J1713.7--3946 exceed this level.
  \label{fig::SignificanceMap}} 
\end{figure}

The background for each grid point was estimated using two different
techniques. The standard \HESS\ background estimation method counts
background events in a ring around each grid point with a mean radius
of 0.5\dg\ and an area 7 times that of the on-region to estimate the
background level. The standard ring radius of 0.5\dg\ was increased to
0.6\dg\ (again a priori) to improve sensitivity to extended
sources. Using appropriate weight factors, this background
determination takes into account the difference in system acceptance
and exposure for the on- and the off-regions~\citep{HESS2155}. An
alternative technique uses the \emph{template} approach, where the
background level is estimated using events that are normally rejected
by the cosmic-ray background rejection criterion based on the scaled
image width and length, but with reconstructed directions falling
within the on-source region~\citep{Template}. The normalisation in
this case is determined by the overall ratio of events in both cut
regimes, taking into account the ratio of relative acceptances in the
two regimes. Figure \ref{fig::Background} shows the correlation of the
background level estimated for each grid point from the ring
background and from the template background method. The slope of the
major axis of the distribution is 1.007. This shows that the
background estimates are consistent within 1\%. The template
background seems to slightly underestimate the background in
comparison to the ring background. The ring background technique
should therefore give a more conservative estimation of the
significances of the new sources. In the following the ring background
estimation method will be used as the standard method, since it is a
robust method in which linear gradients in the system acceptance
cancel out.
From the on-counts, the background level, and the relative background
acceptance in the system, a statistical significance can be determined
for each grid point in the map, testing the existence of a \gr\,
source at the position of that grid point (using the approach of
~\citet{LiMa}). A number of significant sources are detected and
Table~\ref{table::Excess} gives the on-counts, the background level as
derived from the ring background determination method (after
accounting for the difference in acceptance and in the area between
the on-source and the off-source region) and the resulting number of
$\gamma$-ray excess events. Since the map contains a large number of
test positions, this significance has to be corrected according to a
``trial factor''. This trial factor accounts for the increased
probability of finding a fake signal with an increased number of test
positions for which a significance is calculated. Using a Monte-Carlo
simulation of the technique applied to the survey data (including a
realistic exposure map and randomly generating the events for each run
from the acceptance function of that run), it has been found that this
trial factor is somewhat lower than the number of points for which a
significance is calculated, i.e.\ the number of grid points in the map
($n = 220000$), since the excess counts for adjacent grid points are
correlated, given that the integration radius is larger than the grid
spacing. For uncorrelated grid points one would have exactly the
factor $n$. For the calculation of post-trials significances we
conservatively assume that the number of trials is equal to the number
of bins $n$. Figure~\ref{fig::SignificanceDistribution} shows the
distribution of significances for all grid points in the map (solid
black).  The large positive tail to the distribution can be attributed
to the existence of \gr-sources. The much smaller tail of negative
significances is due to the existence of \gr\, sources in the regions
used for background estimation. This effect has been minimised in a
second iteration in which regions of significant \gr\, excess are
excluded as background regions. Also shown in dashed blue is the
distribution of significances for a region in the map without
significant sources ($l = 0 \pm 20^{\circ}$, $0.5^{\circ} < b <
1.5^{\circ}$). This distribution has been scaled to have the same peak
height as the distribution of significances for all grid points
(black). A Gaussian fit to this distribution yields a mean of $-0.050
\pm 0.003$ with an RMS of $1.027 \pm 0.003$, in tolerable agreement
with the expected normal Gaussian distribution given the possibility
of undetected \gr-emission in the on-source or off-source regions.
The distribution of significances for a Monte-Carlo simulation of the
survey without \gr\, sources, using the same analysis procedure as for
data, is shown with red points. This distribution has also been scaled
to have the same peak height as the distribution of significances for
all grid points (black) and is consistent with a normal Gaussian,
demonstrating that the analysis technique behaves as expected.
\begin{table} 
\begin{center}
\caption{Table of on-source counts, scaled background level ($\alpha
  \times N_{\mathrm{off}}$ according to the notation of~\citet{LiMa})
  and $\gamma$-ray excess events for the new sources of \vhe\, \grs\,
  detected in the survey of the Galactic Plane. The on-counts are
  calculated by counting events within a circle of radius $\theta =
  0.22^{\circ}$ centered on the position of the source. The background
  counts are derived by counting events in a ring around the source
  position with average radius 0.6\dg\ multiplied with the
  normalisation factor $\alpha \sim 1/7$ that takes into account the
  difference in the acceptance and the area between the on-source and
  the background region.
\vspace{0.5cm}
\label{table::Excess}} 
  \begin{tabular}{|l|c|c|c|c|}
\tableline\tableline
Name & On-counts & Scaled Background & Excess \\
\tableline
HESS J1614--518& 517 & 283 & 234 \\
HESS J1616--508& 766 & 356 & 410\\
HESS J1632--478& 232 & 139 & 93 \\
HESS J1634--472& 275 & 175 & 100 \\
HESS J1640--465& 611 & 298 & 313 \\
HESS J1702--420& 311 & 210 & 101\\
HESS J1708--410& 955 & 728 & 227 \\
HESS J1713--381& 917 & 728 & 189 \\
HESS J1745--303& 1138 & 927 & 211 \\
HESS J1804--216& 887 & 508 & 379 \\
HESS J1813--178& 717 & 374 & 343 \\
HESS J1825--137& 406 & 232 & 174 \\
HESS J1834--087& 478 & 287 & 191 \\
HESS J1837--069& 622 & 328 & 294 \\
\tableline 
\end{tabular}
\end{center} 
\end{table}
A post-trial significance $S_2$ (the effective \emph{detection
significance}) was calculated from the pre-trials significance $S_1$
by converting $S_1$ into a probability, recalculating the probability
of the occurrence of one or more such events given the number of
trials and converting this probability back into a significance. Grid
points for which the post-trial significance is above 4.0$\sigma$
(corresponding to 6.3$\sigma$ before trials) are considered as likely
sources of \vhe\, \grs\ and are included in the following
discussion. The probability of a single fake signal at this level is
$3\times10^{-5}$ in the whole survey. Figure
\ref{fig::SignificanceMap} shows the map of significances before
accounting for the trial factor with the background estimated from the
ring technique. The strong \gr\, sources RX\,J1713.7--3946 and
HESS\,J1745--290 clearly stand out, partly because of the high
exposure in this regions. Also visible at a much fainter level is the
recently published new \gr\, source G\,0.9+0.1~\citep{HESSG0.9}, which
was serendipitously detected in the \HESS\ observations of the
Galactic Centre region.  In addition to these three known objects,
fourteen new sources of \vhe\, \grs\, are detected at a significance
level above 4$\sigma$ (post-trials), the eight most significant of
which have been discussed in a previous
publication~\citep{HESSScan}. Table \ref{table::Significance} shows
the pre-trials significance for the peak positions in the map for the
new sources in column $S_{1}$ and the significance after accounting
for the number of trials in column $S_{2}$. Column $S_{3}$ shows the
post-trial significance applying a cut appropriate for point sources
($\theta_{\mathrm{point}} =0.1^{\circ}$) instead of the extended
source cut $\theta=0.22\dg$. Only two of the new sources
(HESS\,J1640--465 and HESS\,J1813-178) show a non-marginal increase in
significance using the point-source cut . The extension of the new
sources will be discussed in the next section. It should be noted,
that the five sources with detection significances $S_{2}$ below
5$\sigma$ should be considered as source candidates, rather than
firmly establish sources. Reobservation of these sources will be
necessary to confirm these marginal detections.

\begin{table} 
\begin{center}
\caption{Table of significances for the new sources of \vhe\, \grs\,
  detected in the survey of the Galactic Plane. Source positions are
  derived from a fit to the $\gamma$-ray counts in fine angular bins,
  and do not rely on the coarser spaced search grid. $S_1$ stands for
  the pre-trials significance (using a $\theta^2$ cut of 0.05 deg$^2$)
  for the peak positions in the map. $S_2$ shows the post-trial
  significance for this position. Column $S_{3}$ shows the post-trial
  significance for this position for a point-source cut
  ($\theta_{\mathrm{point}} =0.1^{\circ}$). The sources that are
  marked with a $^\dag$ have been previously reported
  in~\citep{HESSScan}. Galactic longitude $l$ and latitude $b$ as well
  as RA/Dec coordinates are given in degrees, the live time is given
  in hours.
\vspace{0.5cm}
\label{table::Significance}} 
\begin{tabular}{|l|c|c|c|c|c|c|c|c|}
\tableline\tableline
Name & \multicolumn{4}{|c|}{Best Fit Position} &
\multicolumn{3}{|c|}{Significance} & Live time\\
& $l$ ($^{\circ}$)& $b$ ($^{\circ}$)& RA ($^{\circ}$)& Dec($^{\circ}$) & $S_{1}$ & $S_{2}$ & $S_{3}$ &\\
\tableline
HESS J1614--518$^\dag$ & 331.52 & --0.58 & 243.58 & --51.82 & 11.2 & 10.1 & 7.5 & 9.8 \\
HESS J1616--508$^\dag$ & 332.39 & --0.14 & 244.10 & --50.90 & 17.1 & 16.3 & 9.7 & 10.2 \\
HESS J1632--478 & 336.38 & 0.19 & 248.04 & --47.82 & 6.6 & 4.5 & - & 4.5 \\
HESS J1634--472 & 337.11 & 0.22 & 248.74 & --47.27 & 6.4 & 4.2 & 1.8 & 6.6 \\
HESS J1640--465$^\dag$ & 338.32 & --0.02 & 250.18 & --46.53 & 14.5 & 13.6 & 14.5 & 14.3 \\
HESS J1702--420 & 344.26 & --0.22 & 255.69 & --42.07 & 6.3 & 4.0 & 4.1 & 5.7 \\
HESS J1708--410 & 345.67 & --0.44 & 257.06 & --41.08 & 8.4 & 6.8 & 6.9 & 37.4 \\
HESS J1713--381 & 348.65 & 0.38 & 258.49 & --38.20 & 6.3 & 4.0 & 1.4 & 37.3 \\
HESS J1745--303 & 358.71 & --0.64 & 266.26 & --30.37 & 6.3 & 4.0 & 1.8 & 35.3 \\
HESS J1804--216$^\dag$ & 8.40 & --0.03 & 271.13 & --21.70 & 13.9 & 13.0 & 7.7 & 15.7 \\
HESS J1813--178$^\dag$ & 12.81 & --0.03 & 273.40 & --17.84 & 14.4 & 13.5 & 15.3 & 9.7 \\
HESS J1825--137$^\dag$ & 17.82 & --0.74 & 276.51 & --13.76 & 9.5 & 8.1 & 3.7 & 8.4 \\
HESS J1834--087$^\dag$ & 23.24 & --0.32 & 278.69 & --8.76 & 9.5 & 8.1 & 5.7 & 7.3 \\
HESS J1837--069$^\dag$ & 25.18 & --0.11 & 279.41 & --6.95 & 13.2 & 12.2 & 9.9 & 7.6 \\
\tableline 
\end{tabular}
\end{center} 
\end{table}


\section{Characteristics of the new sources}
\label{characteristics}
This section describes the main results and the analysis techniques
that were applied to provide information about position, extension and
spectra of the new \vhe\, \gr\, sources.

\subsection{Position and Morphology}

A detailed spectral and positional analysis has been performed on the
fourteen sources listed in Table~\ref{table::Significance}. For this
purpose, the distribution of $\gamma$-ray candidates was determined
for fine angular bins of width 0.005\dg, to allow for a more accurate
determination of the source position. For the position and approximate
size determination of each source, the spatial distribution of events
in the (uncorrelated) bins was fit to a model of a 2-D Gaussian \gr\,
brightness profile of the form $\rho \propto
\exp(-\theta^{2}/2\sigma_{\mathrm{source}}^2)$, convolved with the PSF
of the instrument, and an added background parametrisation. To
properly handle the small number of events in the bins, a likelihood
fit appropriate for Poisson statistics was applied in this
procedure. To test for possible elongation of each emission region, an
elongated Gaussian with independent $\sigma_{\mathrm{source}}$ and
$\sigma^{\prime}_{\mathrm{source}}$ in each dimension and a free
orientation angle $\omega$ (measured counter-clockwise from the
positive Galactic latitude axis) was also fit. The source is flagged
as elongated if the error range in best fit minor axis
$\sigma^{\prime}_{\mathrm{source}}$ is outside that of the best fit
major axis $\sigma_{\mathrm{source}}$. This fitting procedure has
been tested using Monte-Carlo simulations under realistic
conditions. The re-analysis of the source positions presented here in
comparison to~\citep{HESSScan} yields a maximum shift of 2--3~arc
minutes for the most extended of these sources, reflecting the more
detailed study presented here. The new positions are consistent within
errors with the previously published positions. Also the previously
reported extensions are confirmed by the more detailed analysis
presented here, with the exception of HESS\,J1834-087, where the
extension was overestimated due to the non-Gaussian shape of the
source and is now determined to be 0.09$^{\circ}$ rather than
0.20$^{\circ}$.

\begin{figure}
  \plotone{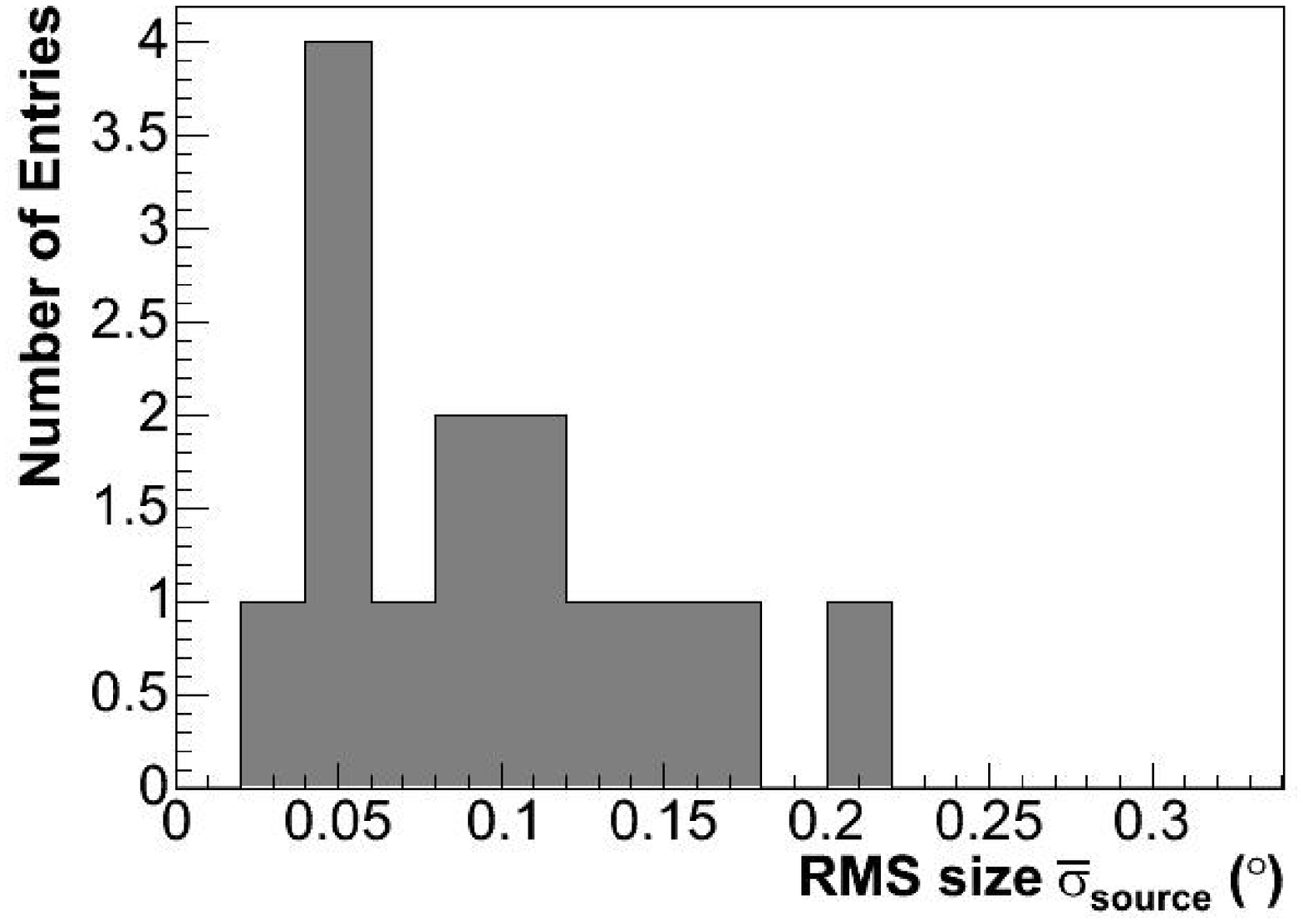}
  \caption{Distribution of the RMS angular size
  $\sigma_{\mathrm{source}}$ of the fourteen \vhe\, \gr\, sources. For
  the elongated sources, the arithmetic mean of
  $\sigma_{\mathrm{source}}$ and $\sigma^{\prime}_{\mathrm{source}}$
  is used. The mean of this distribution is 0.1\dg\ with an RMS of
  0.05\dg.
  \label{fig::SizDistribution}}          
\end{figure}

The system PSF (see Figure~\ref{fig::Offset}) has been derived from
Monte Carlo simulations as a function of the zenith angle of
observations and of the offset of a source in the field of view.  This
model PSF is consistent with the excesses observed for the point-like
Crab Nebula and the active galaxy
PKS\,2155--304~\citep{HESS2155}. Table \ref{table::PositionAndSize}
shows the best fit position, the best fit Gaussian equivalent sizes
$\sigma_{\mathrm{source}}$ and $\sigma^{\prime}_{\mathrm{source}}$,
and the orientation $\omega$ (in case of elongation) of the fourteen
sources. To quantify the quality of the fit, the last column of
Table~\ref{table::PositionAndSize} shows the normalised $\chi^2$
between the model source shape and the measured radial profile fo each
source.
\begin{table} 
\begin{center}
\caption{Position and sizes of the new sources of \vhe\, \grs\,
  detected in the survey of the Galactic Plane. Galactic longitude $l$
  and latitude $b$, the RMS source size $\sigma_{\mathrm{source}}$ and
  $\sigma^{\prime}_{\mathrm{source}}$ and the orientation angle
  $\omega$ counterclock-wise with respect to the positive Galactic
  latitude axis are given in degrees. $\chi^2_{\mathrm{norm}}$ denotes
  the normalised $\chi^2$, with 49 degrees of freedom.
\vspace{0.5cm}
\label{table::PositionAndSize}} 
\begin{tabular}{|l|c|c|c|c|c|c|}
\tableline
\tableline Name & \multicolumn{2}{|c|}{Best Fit Position} &
{$\sigma_{\mathrm{source}}$} & {$\sigma^{\prime}_{\mathrm{source}}$}
 & $\omega$ &{$\chi^2_{\mathrm{norm}}$}\\ & l ($^{\circ}$) & b ($^{\circ}$) & ($^{\circ}$)
& ($^{\circ}$) & ($^{\circ}$) &\\ \tableline 
J1614--518 & 331.52$\pm$0.03 & --0.58$\pm$0.02 & 0.23$\pm$0.02 &
0.15$\pm$0.02 & 49$\pm$10 & 1.44\\
J1616--508 & 332.391$\pm$0.014 & --0.138$\pm$0.013 & 0.136$\pm$0.008 &
- & - & 0.82\\  
J1632--478 & 336.38$\pm$0.04 & 0.19$\pm$0.03 &  0.21$\pm$0.05 &
0.06$\pm$0.04 & 21$\pm$13 & 1.05\\ 
J1634--472 & 337.11$\pm$0.05 & 0.22$\pm$0.04 & 0.11$\pm$0.03 & -
& - & 0.72\\ 
J1640--465 & 338.316$\pm$0.007 & --0.021$\pm$0.007 & 0.045$\pm$0.009 &
- & - & 0.91\\  
J1702--420 & 344.26$\pm$0.04 & --0.22$\pm$0.03& 0.08$\pm$0.04 & -
& - & 1.38\\ 
J1708--410 & 345.672$\pm$0.013& --0.438$\pm$0.010 & 0.054$\pm$0.012 &
- & - & 0.86\\ 
J1713--381 & 348.65$\pm$0.03 & 0.38$\pm$0.03 & 0.06$\pm$0.04 & -
& - & 1.04\\ 
J1745--303 & 358.71$\pm$0.04 & --0.64$\pm$0.05 & 0.21$\pm$0.06 &
0.09$\pm$0.04 & 54$\pm$7 & 1.17\\  
J1804--216 & 8.401 $\pm$0.016 & --0.033$\pm$0.018 & 0.200$\pm$0.010 &
- & - & 1.34\\  
J1813--178& 12.813$\pm$0.005 & --0.0342$\pm$0.005 & 0.036$\pm$0.006 &
- & - & 0.94\\ 
J1825--137& 17.82$\pm$0.03 & --0.74$\pm$0.03 & 0.16$\pm$0.02 & - &
- & 1.03\\ 
J1834--087 & 23.24$\pm$0.02 & --0.32$\pm$0.02 & 0.09$\pm$0.02 & -
& - & 1.16\\ 
J1837--069 & 25.185$\pm$0.012 & --0.106$\pm$0.016 & 0.12$\pm$0.02
& 0.05$\pm$0.02 & 149$\pm$10 & 1.52\\
\tableline 
\end{tabular}
\end{center} 
\end{table}
Figure~\ref{fig::SizDistribution} shows the distribution of sizes for
the new sources. For the elongated sources, the arithmetic mean of
$\sigma_{\mathrm{source}}$ and $\sigma^{\prime}_{\mathrm{source}}$ has
been filled into the histogram. All but two of the sources appear
significantly extended above a level of three standard
deviations. Four out of fourteen appear significantly elongated,
whereas the remaining ten new sources are compatible with a radially
symmetric shape. The radial excess distributions of all sources are
given individually in section~\ref{sec::Sources}.
\subsection{Spectra}

Using the best fit position, a spectral analysis was performed, again
assuming a Gaussian emission region and using a cut of $\theta^2 <
\theta^2_{\mathrm{point}}(\sigma_{\mathrm{source}}^{2}+\sigma_{\mathrm{PSF}}^{2})/\sigma_{\mathrm{PSF}}^{2}$,
approximating the \HESS\ PSF as a 2-D Gaussian with
$\sigma_{\mathrm{PSF}}$ = 0.08\dg\ and $\theta_{\mathrm{point}}
=0.1^{\circ}$ being the optimal cut for point sources. Such a cut
provides optimum significance for a weak Gaussian shaped source on a
uniform background, and hence the best possible spectral
determination. For the elongated sources, the larger of the two values
$\sigma_{\mathrm{source}}$ and $\sigma^{\prime}_{\mathrm{source}}$ was
used. The energy dependent efficiency of this $\theta$ cut is
estimated by convolving the \HESS\ PSF with a Gaussian of size
$\sigma_{\mathrm{source}}$. This introduces an additional systematic
error in the absolute flux determination since it implicitly assumes
that the emission region of the source follows a Gaussian distribution
and that the morphology of the source does not show any energy
dependency. The typical overall systematic error in the determination
of the integral flux level is estimated to be $\sim30$\%. The standard
\HESS\ spectral analysis technique is described
in~\citep{HESS2155}. To avoid systematic errors in the spectral
determination arising from different camera acceptances across the
H.E.S.S. field of view, the spectral background determination employs
regions of the same size and shape as the \textit{on} region, but
displaced on a ring in the field of view around the camera centre. The
ring was chosen such that the \textit{off} regions have the same
angular distance to the centre as the \textit{on} region. This
approach ensures that background events are taken at the same zenith
and offset angles.

\begin{figure}
  \plotone{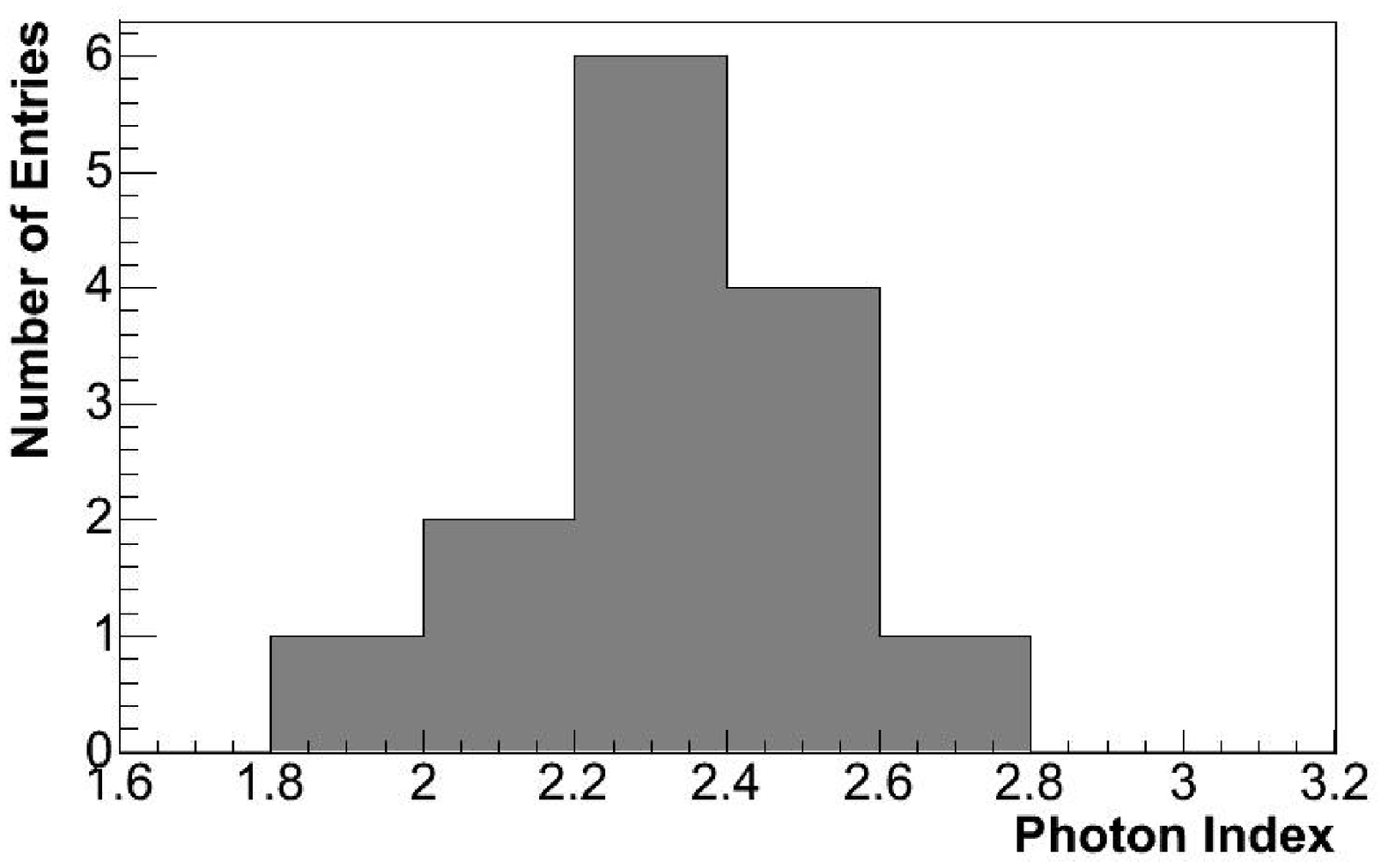}
  \caption{Distributions of the photon index of the new sources. The
  mean photon index is 2.32 with an RMS of 0.2. 
  \label{fig::IndexDistribution}}          
\end{figure}

To keep as many low-energy \grs\, as possible, a looser size cut of
80~photo-electrons is used in the spectral analysis. This cut allows
to reach a lower energy threshold (150~GeV) at the expense of reduced
sensitivity for hard-spectrum sources and a 30\% poorer angular
resolution. Flux points below a significance level of 2 standard
deviations were discarded in the spectral analysis and the binning was
doubled at high energies for weak sources. Table~\ref{table::Spectra}
lists the resulting photon index $\Gamma$ and flux above 200~GeV for a
power-law fit $dN/dE = F_{0} E^{-\Gamma}$. The typical systematic
error on these photon indices is estimated to be $\sim0.2$. It should
be noted, that the fluxes reported here are in general different from
the fluxes reported in~\citet{HESSScan} (especially for the larger
sources) since in the present work the flux is determined from a
larger region, adapted to encompass a larger fraction of the emission
region, whereas in~\citet{HESSScan}, a fixed small $\theta^2=0.02$
deg$^2$ cut was used for all sources.
\begin{table} 
\begin{center}
\caption{Results of a power-law fit $dN/dE = F_{0} E^{-\Gamma}$ for
  the fourteen sources of \vhe\, \grs.
  The $\theta$ cut used for the spectral analysis is given in
  degrees. The errors given on the photon index and flux are
  statistical only. The typical systematic errors are $\sim0.2$ in
  the photon index and $\sim30$\% in the absolute flux. To derive a
  combined error on the photon index, the systematic and statistical
  errors should be added in quadrature.
\vspace{0.5cm}
\label{table::Spectra}} 
\begin{tabular}{|l|c|c|c|c|}
\tableline\tableline
Name & $\Gamma$ & Flux $> 200$ GeV & $\theta$ cut & $\chi^2$/NDF \\
& & ($10^{-12}$ cm$^{-2}$ s$^{-1}$)& ($^{\circ}$)&\\
\tableline
HESS\,J1614--518 & 2.46$\pm$0.20 & 57.8$\pm$7.7 & 0.40 & 0.6/3 \\
HESS\,J1616--508 & 2.35$\pm$0.06 & 43.3$\pm$2.0 & 0.26 & 5.5/5\\
HESS\,J1632--478 & 2.12$\pm$0.20 & 28.7$\pm$5.3 & 0.36 & 7.9/2 \\
HESS\,J1634--472 & 2.38$\pm$0.27 & 13.4$\pm$2.6 & 0.23 & 5.8/5\\
HESS\,J1640--465 & 2.42$\pm$0.15 & 20.9$\pm$2.2 & 0.16 & 3.7/4\\
HESS\,J1702--420 & 2.31$\pm$0.15 & 15.9$\pm$1.8 & 0.20 & 7.8/6 \\
HESS\,J1708--410 & 2.34$\pm$0.11 & 8.8$\pm$0.7 & 0.17 & 6.3/3 \\
HESS\,J1713--381 & 2.27$\pm$0.48 & 4.2$\pm$1.5 & 0.17 & 1.8/2 \\
HESS\,J1745--303 & 1.82$\pm$0.29 & 11.2$\pm$4.0 & 0.36 & 4.4/3 \\
HESS\,J1804--216 & 2.72$\pm$0.06 & 53.2$\pm$2.0 & 0.36 & 6.9/11 \\
HESS\,J1813--178 & 2.09$\pm$0.08 & 14.2$\pm$1.1 & 0.15 & 5.5/11 \\
HESS\,J1825--137 & 2.46$\pm$0.08 & 39.4$\pm$2.2 & 0.30 & 6.2/8 \\
HESS\,J1834--087 & 2.45$\pm$0.16 & 18.7$\pm$2.0 & 0.20 & 3.5/4 \\
HESS\,J1837--069 & 2.27$\pm$0.06 & 30.4$\pm$1.6 & 0.23 & 12.6/10 \\
\tableline 
\end{tabular}
\end{center} 
\end{table}
Figure~\ref{fig::IndexDistribution} shows the distribution of photon
indices for the fourteen new sources. The distribution exhibits a mean
of 2.32, marginally consistent with predictions by models of shock
wave acceleration in SNRs (see e.g.~\citet{BerezhkoVoelk1997,
BerezhkoVoelk2000}) and as expected for the source spectrum of
Galactic cosmic rays (see e.g~\citet{Jones}). The distribution has an
RMS of 0.2, comparable with the statistical errors on the
measurements.
\subsection{Individual \gr\, sources}
\label{sec::Sources}

This section describes the fourteen \vhe\, \gr\, sources detected in
the Galactic Plane survey individually along with a brief description
of possible counterparts at other wavelengths. Excess maps are shown
that have been smoothed with a two-dimensional Gaussian to reduce the
effect of statistical fluctuations. The resulting count map is in
units of integrated excess counts within the smoothing radius. The RMS
radius of the Gaussian used for the smoothing is adapted for each
source according to the available photon statistics. This radius is
given in the figure captions, and is shown as a black dotted line in
the inset in the lower right part of each excess map along with the
PSF of the instrument, smoothed in the same way as the excess
map. Note that the maps have different scales, therefore the white
line in the upper right corner indicates an 0.2\dg\ scale. Further it
should be noted that the colour scale covers different intervals for
different maps. 

Also shown are possible counterparts to the \vhe\, \gr\, emission: SNR
from~\citet{Green} (with white circles illustrating their nominal
radii), pulsars from the ATNF pulsar catalogue~\citep{ATNF} (as white
triangles), hard X-ray sources detected by the International Gamma-Ray
Laboratory (INTEGRAL) satellite (as white stars), and unidentified
EGRET sources from the third EGRET catalogue~\citep{EGRET} (showing
the 95\% error contour for the position of the source in dashed
white). In some maps contours of radio and/or X-ray emission are also
shown and discussed individually. 20~cm Very large array (VLA) radio
contours are taken from~\citet{RadioGalacticPlane}. The x-ray ASCA
data show smoothed count maps obtained with the GIS instrument in the
0.7--10~keV band. The best fit source position of the H.E.S.S. source
is shown with error bars in black, the best fit (Gaussian) source size
is shown as a black dashed circle or ellipse. The radial distribution
of \gr\, excess with respect to the best-fit source position is shown
for each source binned appropriately for the available photon
statistics; in each plot the number of reconstructed events in bins of
the squared angular distance ($\theta^2$) is shown along with the
best-fit radially symmetric model (solid black line) and the PSF of
the instrument (dashed red line), derived from Monte-Carlo
simulations. For all sources the reconstructed differential energy
spectra are shown together with the best-fit power-law (solid line).
 
\subsubsection{HESS\,J1614--518}
\begin{figure}
  \plotone{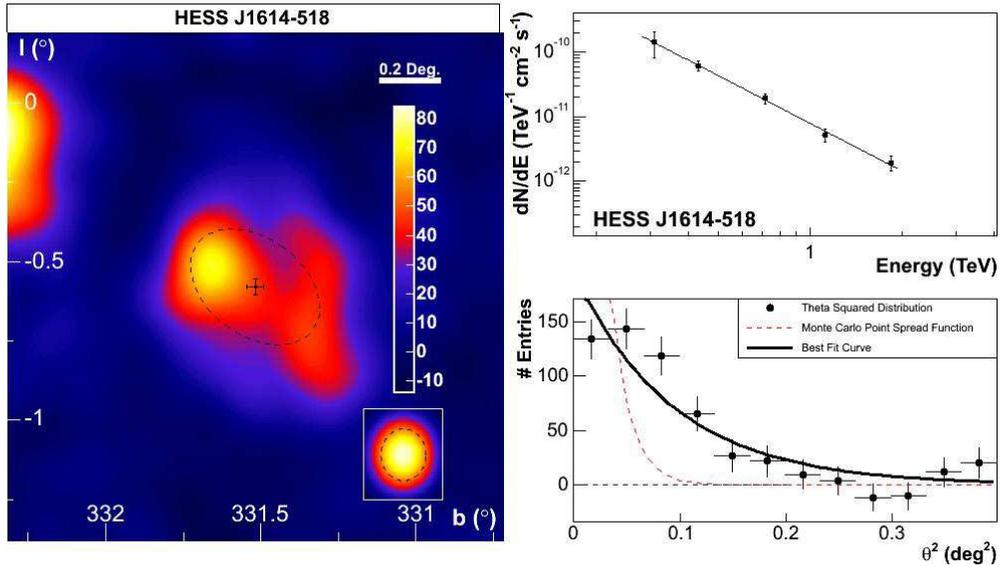}
  \caption{Smoothed excess map (left, smoothing radius 0.09\dg) of the
  region of HESS\,J1614--518. The figures on the right hand side show
  the energy spectrum (top) and $\theta^2$ plot (bottom).
  \label{fig::Hotspot9}}          
\end{figure}

This source (see Figure~\ref{fig::Hotspot9}) is located in a region
relatively devoid of counterpart candidates. The source exhibits
elliptical emission with a semi major axis of 14$\pm$1~arc minutes and
a semi minor axis of 9$\pm$1~arc minutes. HESS\,J1614--518 is the
brightest of the new sources, detected with a flux corresponding to
25\% of that from the Crab Nebula above 200~GeV. The spectrum is well
fit by a power-law of photon index 2.46$\pm$0.20, see
Figure~\ref{fig::Hotspot9}~(top right). A nearby \emph{Chandra}
observation~\citep{ChandraMz3} covered an area more than 10~arc
minutes away from the best fit position of HESS\,J1614--518 and showed
no evidence for X-ray emission from the direction of the
H.E.S.S. source. The poor quality of the fit to the radial
distribution of HESS\,J1614--518 as is shown in
Figure~\ref{fig::Hotspot9}~(bottom right) indicates that the source is
not well described by a symmetrical Gaussian profile.
\subsubsection{HESS\,J1616--508}
\begin{figure}
  \plotone{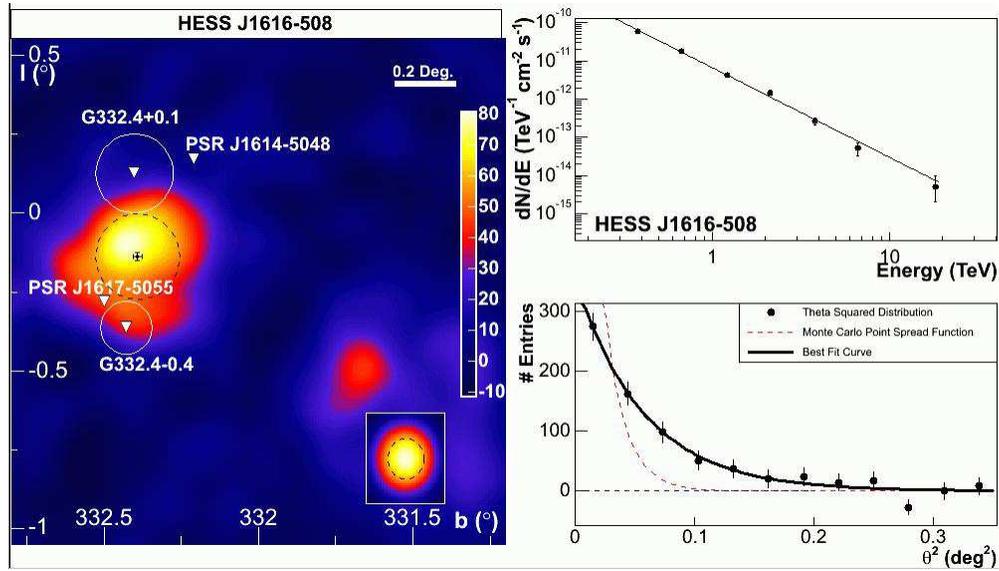}
  \caption{Smoothed excess map (left, smoothing radius 0.06\dg) of the
  region surrounding HESS\,J1616--508 (left hand source), along with
  nearby pulsars and SNRs which were considered as
  counterparts. HESS\,J1614--518 is also visible in this map (right
  hand source). The figures on the right hand side show the energy
  spectrum (top) and the radial distribution ($\theta^2$) plot
  (bottom).
  \label{fig::Hotspot3}}          
\end{figure}

HESS\,J1616--508 (see Figure~\ref{fig::Hotspot3}) with a flux of 19\%
of the Crab Nebula above 200~GeV is also one of the brighter new
sources of \vhe\, \grs. Its spectrum can be fitted by a power-law of
photon index 2.35$\pm$0.06 as shown in Figure~\ref{fig::Hotspot3}~(top
right). The emission region has an RMS size of 16~arc minutes and does
not have a good positional match to any plausible counterpart at other
wavelengths. It is located in a region close to the known SNRs
G332.4--0.4 (RCW\,103) and G332.4+0.1 (Kes\,32). The region has been
intensively studied with X-ray satellites like
\emph{Chandra}~\citep{G332.4+0.1Chandra} and \emph{XMM}, with the
position of HESS\,J1616--508 mostly outside or at the edge of the
field of view where the sensitivity is strongly decreased. No extended
X-ray source that lines up with the \vhe\, \gr\, emission has been
found. Another interesting object in this region is the energetic
pulsar PSR\,J1617--5055~\citep{PSRJ1617Detection}, a young
X-ray-emitting pulsar, that was proposed to emit TeV
\grs~\citep{Pulsar1617TeV}. Approximately 1.2\% of its spin-down
luminosity would suffice to power the observed (0.2--10 TeV)\vhe\,
\gr\, flux, assuming the H.E.S.S. source is located at the pulsar
distance. Since HESS\,J1616--508 does not line up with this object, an
asymmetric (yet undetected) PWN nebula powered by the pulsar would be
required to explain the TeV emission.
\subsubsection{HESS\,J1632--478}

\begin{figure}
  \epsscale{.80}
  \plotone{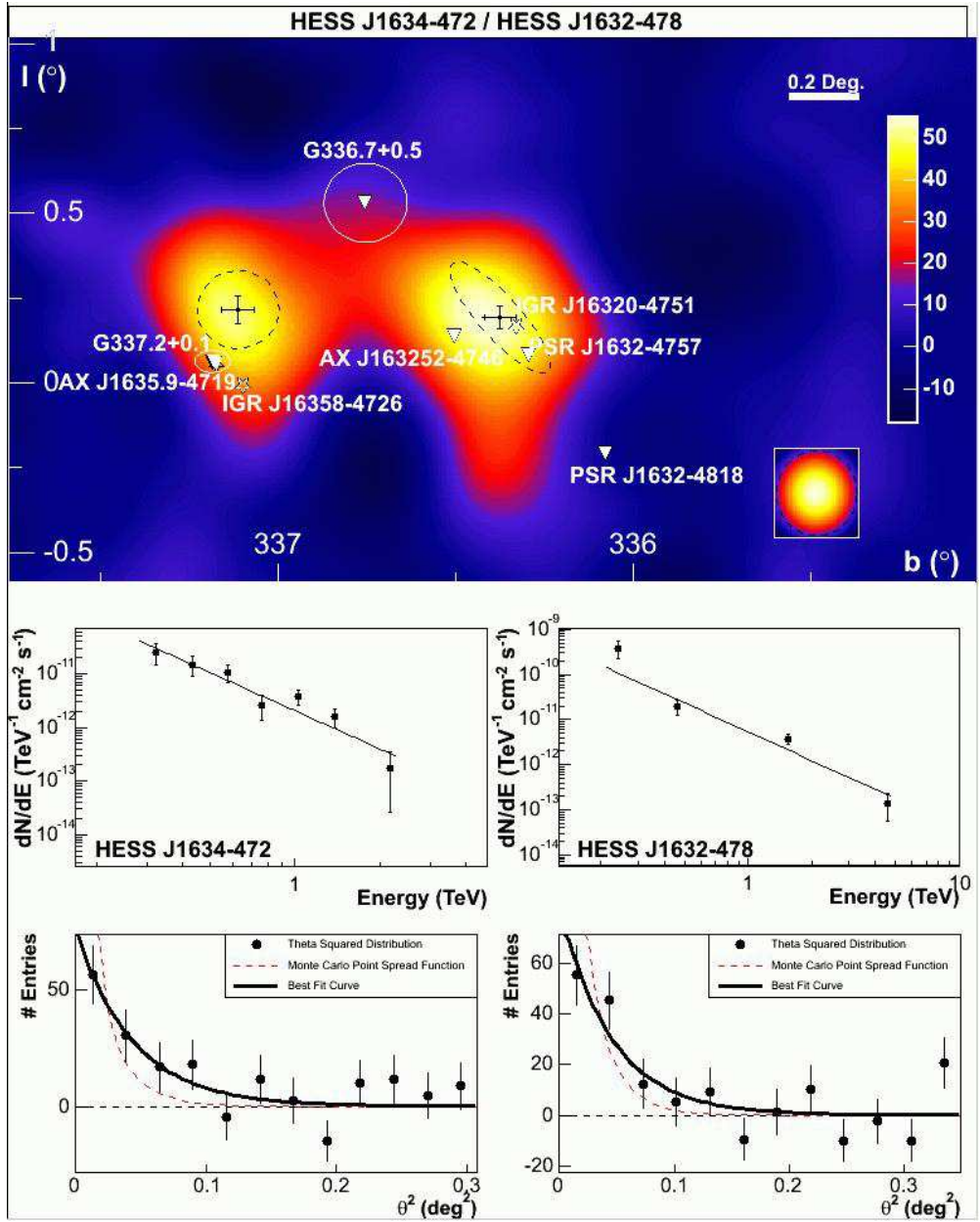}
  \caption{Smoothed excess map (top, smoothing radius 0.12\dg) of the
  region of the two close-by \HESS\ sources HESS\,J1634--472 (left
  hand source) and HESS\,J1632--478 (right hand source), along with
  nearby pulsars, SNRs, and INTEGRAL sources which were considered as
  counterparts. It should be noted, that the tail of HESS\,J1632--478
  to negative Galactic latitude is statistically not significant. The
  figures in the middle row show the spectra (left: HESS\,J1634--472,
  right HESS\,J1632--478), the figures in the bottom row the
  corresponding $\theta^2$ plots.
    \label{fig::Hotspot15}}          
\end{figure}

The centroid position of this object (right hand source in
Figure~\ref{fig::Hotspot15}~(top)) and of the close-by
HESS\,J1634--472 are spatially separated by less than 45~arc
minutes. HESS\,J1632--478 has an elongated shape with a semi-major
axis of 12$\pm$3~arc minutes and a semi-minor axis of 3.6$\pm$2.4~arc
minutes. Its flux in \vhe\, \grs\, above 200~GeV corresponds to 12\%
of the flux from the Crab Nebula and the spectrum can be fitted by a
power-law of photon index 2.12$\pm$0.20 (see
Figure~\ref{fig::Hotspot15}~(bottom right)). A positional coincidence
of HESS\,J1632--478 exists with the hard X-ray source
IGR\,J16320--4751 discovered by the INTEGRAL
satellite~\citep{INTEGRAL16320} in the energy range above 15~keV. In
softer X-rays between 2 and 10~keV, this object was also strongly
detected in an \emph{XMM-Newton} observation~\citep{XMM16320} and
found to be coincident with the ASCA source AX\,J1631.9--4752,
detected in the ASCA Galactic Plane survey~\citep{ASCA}. This source
is believed to belong to a new class of heavily absorbed high mass
X-ray binary systems, with an equivalent absorption column density of
$(9.2 \pm 1.1) \times 10^{22}$~cm$^{-2}$ in this case. The extension
of HESS\,J1632--478 does not plead in favor of an association with
IGR\,J16320--4751. One should note that the probability of spurious
associations is increased in this region of the Galactic plane (the
3~kpc arm tangent region) due to the high density of absorbed INTEGRAL
sources. Another potential counterpart is the ASCA source
AX\,J163252--4746.

\subsubsection{HESS\,J1634--472}
The size of the HESS\,J1634--472 (left hand source in
Figure~\ref{fig::Hotspot15}~(top)) emission region is 6.6$\pm$1.8~arc
minutes and its flux above 200~GeV is close to 6\% of the Crab flux
with a photon index for the energy spectrum of 2.38$\pm$0.26 (see
Figure~\ref{fig::Hotspot15}, centre left). HESS\,J1634--472 has no
direct positional counterpart, although it is interesting to note that
yet another INTEGRAL source IGR\,J16358--4726 is located
close-by~\citep{INTEGRAL16358}. This source presumably belongs to the
class of highly absorbed X-ray binary systems, like IGR\,J16320--4751
(discussed in the previous paragraph). The source was detected in the
2--10~keV X-ray band with the \emph{Chandra}
satellite~\citep{Chandra16358}. The column density dermined for this
object is $N_{H}= 3.3 \times 10^{23}$~cm$^{-2}$ and the luminosity of
this object in the 2--10~keV band is $10^{36}$~erg/s. Another
potential counterpart of HESS\,J1634--472 is the recently discovered
SNR candidate G337.2+0.1~\citep{SNRG337.2}, coincident with the ASCA
source AX\,J1635.9--4719. However, the distances of the two objects to
HESS\,J1634--472 renders a correlation unlikely.

\subsubsection{HESS\,J1640--465}
\begin{figure}
  \plotone{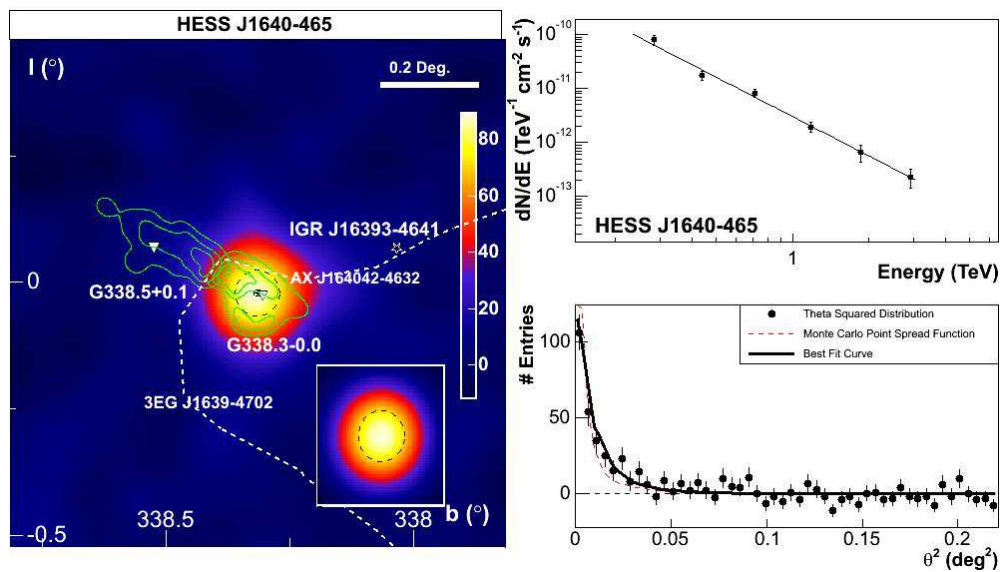}
  \caption{Smoothed excess map (left, smoothing radius 0.05\dg) of the
  region surrounding HESS\,J1640--465 along with nearby SNRs which
  were considered as counterparts as discussed in the text. The green
  contours indicate the radio emission detected from the region
  surrounding G338.3--0.0 in the Molonglo Galactic Plane survey at
  843~MHz~\citep{Molonglo}. The dashed white contours show the 95\%
  positional error of the unidentified EGRET source
  3EG\,J1639-4702. The differential energy spectrum (top) and the
  radial distribution ($\theta^2$) plot (bottom) are shown on the
  right hand side.
  \label{fig::Hotspot4}}          
\end{figure}

As can be seen from the radial distribution of this source (see
Figure~\ref{fig::Hotspot4}~(bottom right)) HESS\,J1640--465 is
marginally extended with respect to the PSF of the instrument. It is
one of the few examples of a excellent spatial correlation between the
\vhe\, \gr\, emission and an SNR (G338.3--0.0), known from radio
observations~\citep{Green}. G338.3--0.0 was observed in the Molonglo
Galactic Plane survey at 843~MHz (shown in green contours in
Figure~\ref{fig::Hotspot4}~(left)) and has been reported to be a
broken shell SNR lying on the edge of a bright HII
region~\citep{Whiteoak}. The region bridging this SNR and the close-by
SNR G\,338.5+0.1 is known to contain dense HII regions. The relation
of surface-brightness to distance yields a distance of 8.6~kpc, which
would place the SNR in the 3~kpc arm tangent region. The ASCA source
AX\,J164042-4632 with a photon index of $2.98^{+1.13}_{-0.89}$ and a
flux level of $1.21\times 10^{-12}$ ergs cm$^{-2}$ s$^{-1}$, detected
during the ASCA Galactic Plane survey~\citep{ASCA}, has been
identified with G338.3--0.0. It should be noted, that the position of
HESS\,J1640--465 is compatible with the position of the unidentified
EGRET source 3EG\,J1639--4702~\citep{EGRET}. The 95\% positional
confidence contour of 3EG\,J1639--4702 is shown in
Figure~\ref{fig::Hotspot4}~(left). The distance of HESS\,J1640--465 to
this EGRET source is 34~arc minutes. The differential energy spectrum
of HESS\,J1640--465 can be well fit by a power-law with photon index
2.42$\pm$0.14 at a flux close to 9\% of that from the Crab Nebula
above 200~GeV (see Figure~\ref{fig::Hotspot4}~(top right)). A simple
power-law extrapolation of the VHE gamma-ray spectrum to 1~GeV matches
the EGRET flux above that energy.
\subsubsection{HESS\,J1702--420}
\begin{figure}
  \plotone{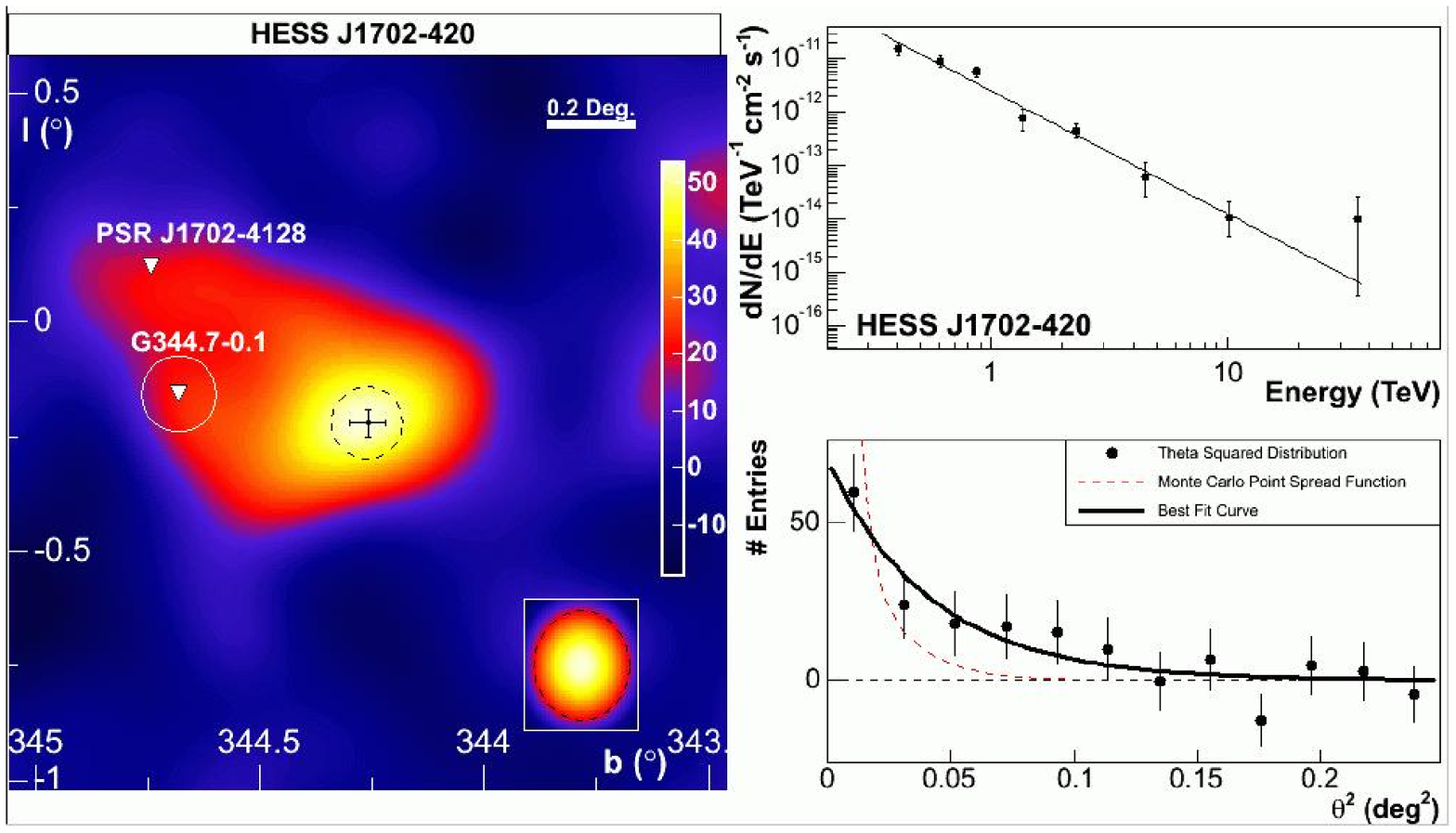}
  \caption{Smoothed excess map (left, smoothing radius 0.10\dg) of the
  region surrounding HESS\,1702--420 along with nearby pulsars and
  SNRs which were considered as counterparts as discussed in the
  text. The tail extending to positive Galactic longitude and latitude
  is not statistically significant. The differential energy spectrum
  (top) and the radial distribution ($\theta^2$) plot (bottom) are
  shown on the right hand side.
  \label{fig::Hotspot11}}          
\end{figure}

For HESS\,J1702--420 (Figure~\ref{fig::Hotspot11}) no significant
evidence for extension of the emission region can be derived beyond a
level of 2 standard deviations for the measured source size of
4.8$\pm$2.4~arc minutes. The flux above 200~GeV of this source
corresponds to 7\% of the Crab flux and the energy spectrum can be
fitted by a power-law of photon index 2.31$\pm$0.15 (see
Figure~\ref{fig::Hotspot11}~(top right)). The close-by pulsar
PSR\,J1702--4128 is energetic enough to account for the observed
\vhe\, \gr\, emission by converting roughly 14\% of its spin-down
luminosity but would need an asymmetric PWN to be responsible for the
observed \vhe\ $\gamma$-ray emission. No asymmetric PWN has been
detected from this pulsar, however, which renders an association
unlikely. The nearby SNR G\,344.7--0.1 is too small and not
positionally coincident with the \vhe\ \gr-source. Therefore, we
cannot identify a plausible counterpart for this object.
\subsubsection{HESS\,J1708--410}
\begin{figure}
  \plotone{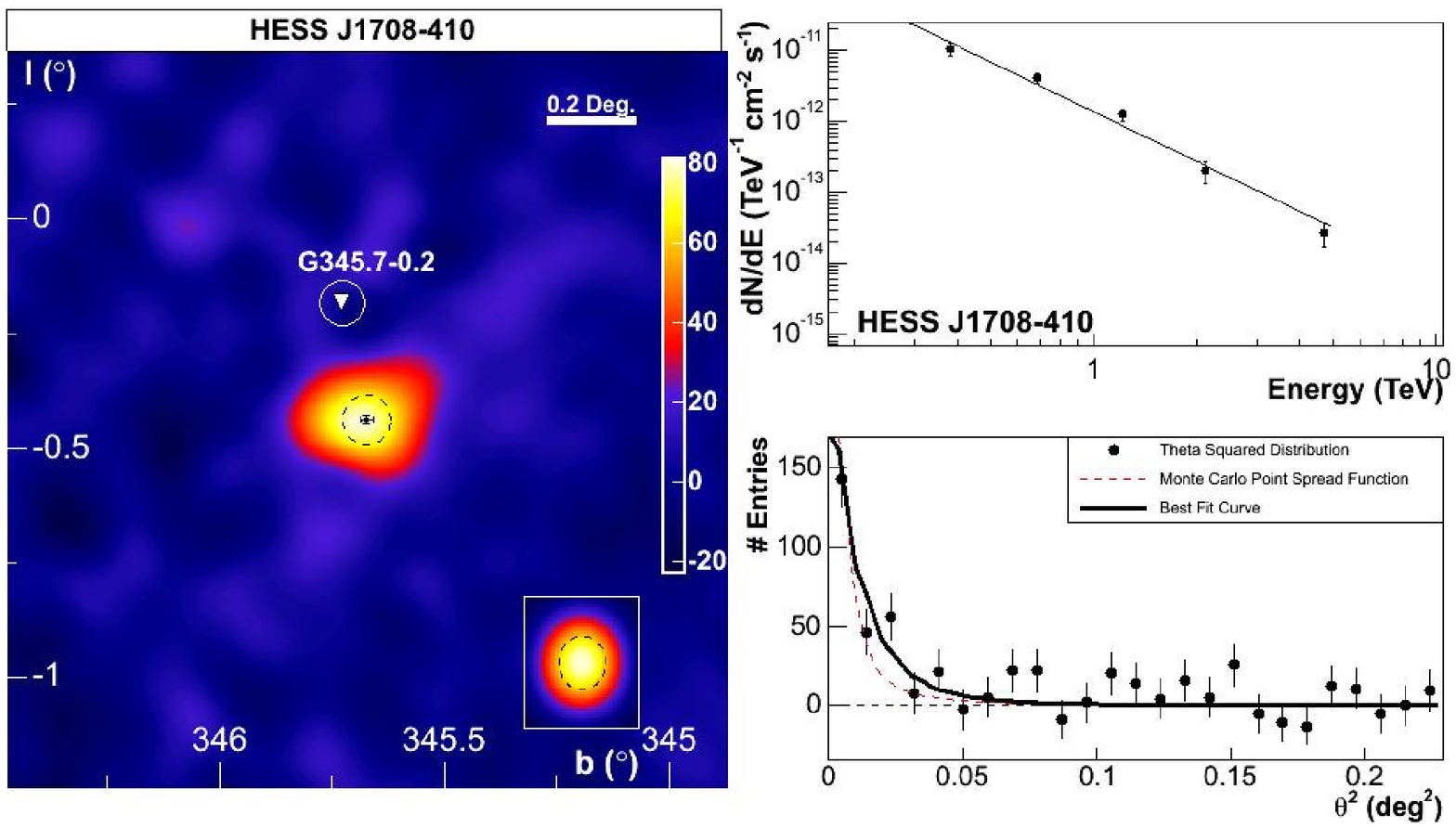}
  \caption{Smoothed excess map (left, smoothing radius 0.05\dg) of the
    region surrounding HESS\,1708--410 along with a nearby SNR which
    was considered as counterpart. The differential energy spectrum
    (top) and the radial distribution ($\theta^2$) plot (bottom) are
    shown on the right hand side.
  \label{fig::Hotspot10}}          
\end{figure}

This \vhe\, \gr\, source (Figure~\ref{fig::Hotspot10}) of size
3.2$\pm$0.7~arc minutes is located very close to SNR RX\,J1713.7--3946
and therefore its observation live time (within 2\dg\ of the pointing
position of the instrument) amounts to 37.2~hours, well beyond typical
exposures in the \HESS\ survey. The source exhibits an energy spectrum
with a power-law index of 2.34$\pm$0.11, the flux above 200~GeV is
less than 4\% of the flux from the Crab Nebula (see
Figure~\ref{fig::Hotspot10}~(top right)). As can be seen from
Figure~\ref{fig::Hotspot10}~(left), there is no close-by plausible
counterpart. A nearby \emph{XMM} observation of the SNR G345.7--0.2
has revealed no X-ray counterpart to the \vhe\, \gr\, source, but it
should be noted that the offset of HESS\,J1708--410 from the centre of
the \emph{XMM} field of view was 15~arc minutes, positioning it close
to the edge of the field of view. As can be seen from the radial
distribution, shown in Figure~\ref{fig::Hotspot10}~(bottom right), the
source is only marginally extended with respect to the PSF of the
instrument.
\subsubsection{HESS\,J1713--381}
\begin{figure}
  \plotone{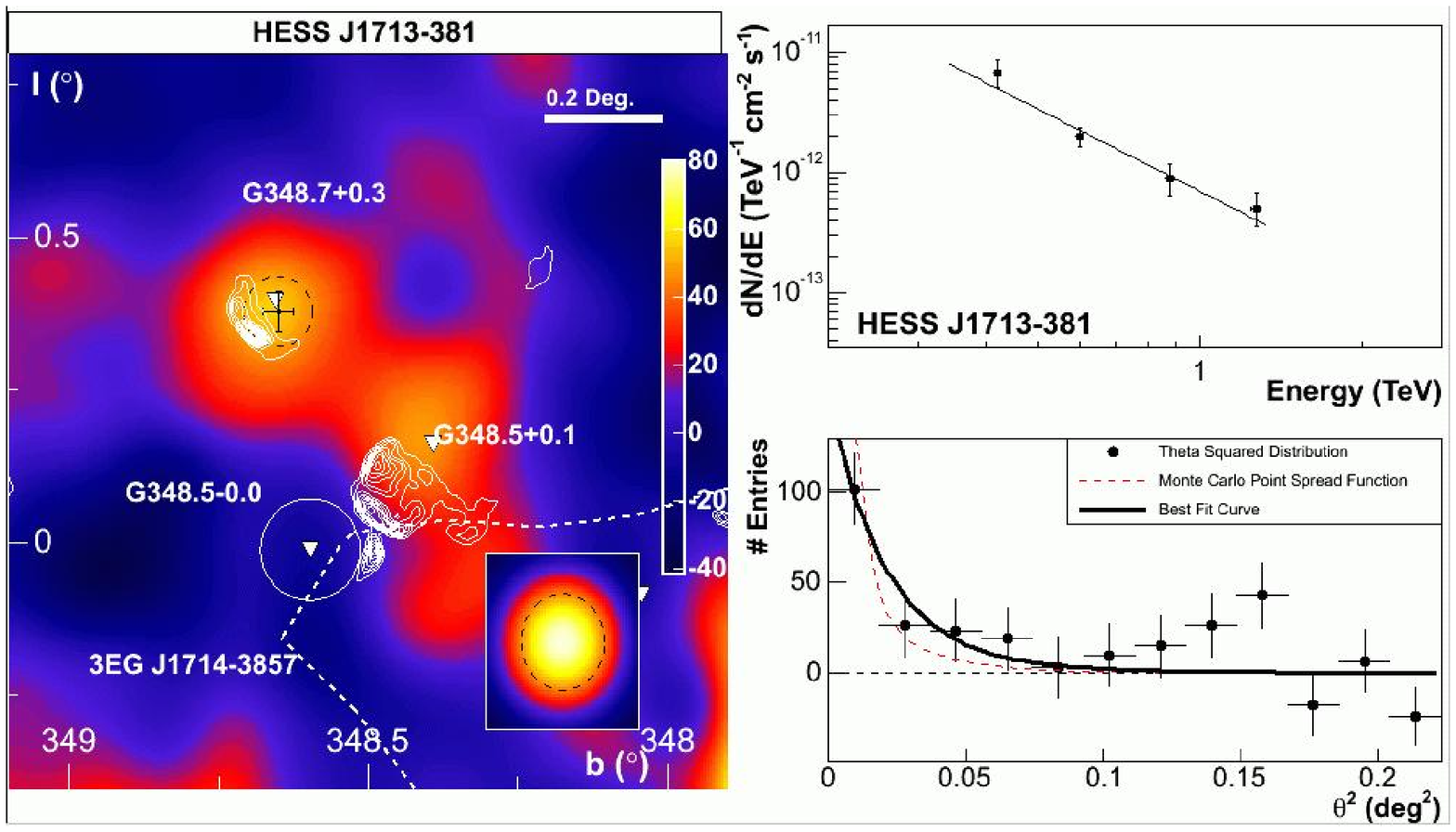}
  \caption{Smoothed excess map (left, smoothing radius 0.07\dg) of the
  region surrounding HESS\,J1713--381 along with nearby SNRs which
  were considered as counterparts as discussed in the text. In the
  bottom right corner of this excess map, the strong \vhe\, emission
  from SNR RX\,J1713.7--3946 is apparent. The white contours show the
  radio emission as seen in the Molonglo Plane survey at
  843~MHz~\citep{Molonglo}. The marginal excess in positional
  agreement with G\,348.5+0.1 (CTB~37\,A) is not significant after
  accounting for all trials for the analysis presented here. The
  differential energy spectrum (top) and the radial distribution
  ($\theta^2$) plot (bottom) are shown on the right hand side. Also in
  the radial distribution, some indication for an excess $\sim0.4$\dg\
  away from HESS\,J1713--381 can be seen, due to the marginal excess
  at the position of CTB~37\,A. The dashed white contours show the
  95\% positional error of the unidentified EGRET source
  3EG\,J1714--3857.
  \label{fig::Hotspot12}}          
\end{figure}

This source (Figure~\ref{fig::Hotspot12}) is the second of the
fourteen \vhe\, sources that shows no significant evidence beyond 2
standard deviations for an extension of the emission region. The
energy spectrum of HESS\,J1713--381 is consistent with a power-law
with photon index 2.2$\pm$0.5. Its flux of 1.8\% of the flux from the
Crab Nebula above 200~GeV makes this object the weakest of the new
sources. Its location is very close to the strong \vhe\,\gr\, source
SNR RX\,J1713.7--3946 and it was therefore in the field of view of
most pointed observations by \HESS\ on this object. The
correspondingly deep exposure of 37~hours at offsets of less than
2$^{\circ}$ from the centre of the field of view made it possible to
detect this object. As can be seen from
Figure~\ref{fig::Hotspot12}~(left), it coincides with part of the
unusual SNR complex CTB~37. The radio contours of the CTB~37 complex
as taken from the Molonglo Plane survey at 843~MHz are shown in
white~\citep{Molonglo}. There is clear positional match between the
source HESS\,J1713--381 and the SNR G\,348.7+0.3 (CTB~37\,B). Also
visible in Figure~\ref{fig::Hotspot12}~(right) is a weaker marginal
source consistent with the position of G\,348.5+0.1 (CTB~37\,A). The
marginal detection of emission from this source is however not
significant after accounting for all trials for the analysis of
extended sources in this survey as described above. ~\citet{KassimCTB}
describe CTB~37\,A as two SNRs overlapping in projection: with the
partial shell of G\,348.7+0.1 in the west and the SNR G\,348.5--0.0 to
the east. Both SNRs in CTB\,37A appear to be interacting with a system
of molecular clouds with densities of
100--1000~cm$^{-3}$~\citep{COCTB}.  The SNR CTB~37\,B is less well
studied but appears to share the dense environment of
CTB~37\,A. Evidence for X-ray emission from the CTB~37 complex can
also be seen in archival ASCA data.
\subsubsection{HESS\,J1745--303}
\begin{figure}
  \plotone{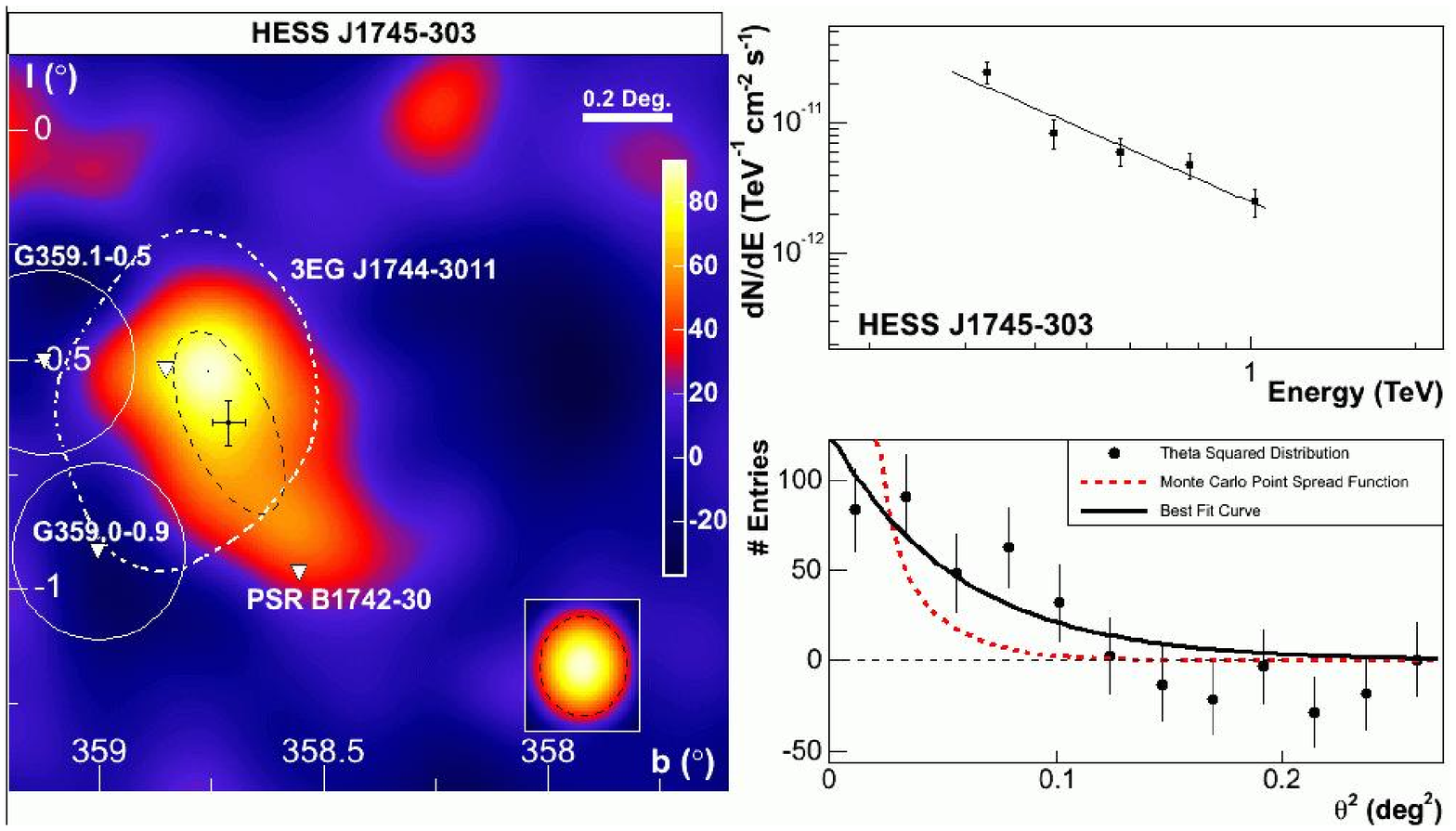}
  \caption{Smoothed excess map (left, smoothing radius 0.10\dg) of the
    region surrounding HESS\,J1745--303 along with nearby pulsars and
    SNRs which were considered as counterparts as discussed in the
    text. The differential energy spectrum (top) and the radial
    distribution ($\theta^2$) plot (bottom) are shown on the right
    hand side.
  \label{fig::Hotspot14}}          
\end{figure}

This object (Figure~\ref{fig::Hotspot14}) has a direct positional
coincidence with an unidentified EGRET source
(3EG\,J1744--3011)~\citep{EGRET}. Since it is within the field of view
of the Galactic Centre, HESS\,J1745--303 has a large observation live
time of 35.3~hours. Its flux level corresponds to 5\% of the flux from
the Crab Nebula above 200~GeV and the emission exhibits a hard
differential flux of index 1.8$\pm$0.3. Its emission region has an
elliptical shape with a semi-major axis of 13$\pm$4~arc minutes and a
semi-minor axis of 5.4$\pm$2.4~arc minutes as is shown in
Figure~\ref{fig::Hotspot14}~(left). The distance of HESS\,J1745--303
to the centroid of the EGRET source 3EG\,J1744--3011 is 10~arc
minutes, which is well within the 95\% uncertainty level of the EGRET
position. A simple power-law extrapolation of the \vhe\, \gr\,
spectrum to energies above 1~GeV is an order of magnitude below the
EGRET flux above that energy ($F(>100\ \mathrm{MeV}) = (64 \pm
7)\times 10^{-8}$ cm$^{-2}$ s$^-1$). Another potential counterpart is
the pulsar PSR\,B1742--30 but essentially 100\% of its spin-down
power would be required to account for the entire \vhe\ $\gamma$-ray
emission.
\subsubsection{HESS\,J1804--216}
\begin{figure}
  \plotone{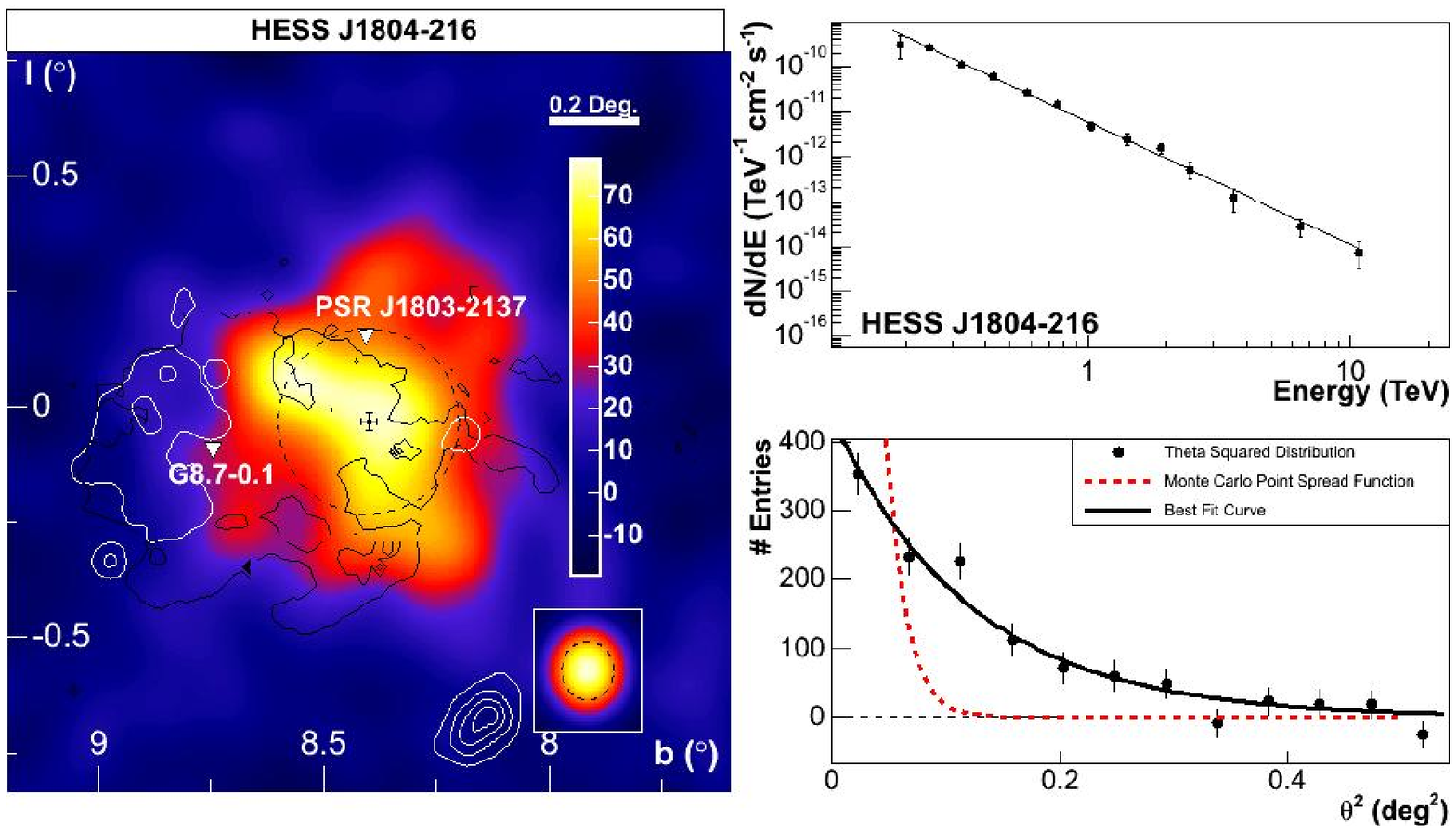}
  \caption{Smoothed excess map (left, smoothing radius 0.06\dg) of the
    region surrounding HESS\,J1804--216 along with nearby pulsars and
    SNRs which were considered as counterparts as discussed in the
    text. The white contours show the X-ray emission from G8.7--0.1 as
    detected by the ROSAT satellite, the black contours show the 20~cm
    radio emission detected by the VLA. The differential energy
    spectrum (top) and the radial distribution ($\theta^2$) plot
    (bottom) are shown on the right hand side.
  \label{fig::Hotspot2}}          
\end{figure}

This source (Figure~\ref{fig::Hotspot2}) with a size of approximately
22~arc minutes is the largest of the new emission regions of \vhe\,
\grs. With a flux of nearly 25\% of the flux from the Crab Nebula
above 200~GeV, it is also the brightest of the new \HESS\ sources. The
photon index of 2.72$\pm$0.06 makes it one of the softest sources (see
Figure~\ref{fig::Hotspot2}~(top right)). As can be seen from
Figure~\ref{fig::Hotspot2}~(left), the emission region does not
perfectly line up with any known object, although it could be
associated with the south-western part of the shell of the SNR
G8.7--0.1 of radius 26~arc minutes. From CO observations, the
surrounding region is known to be associated with molecular gas where
massive star formation takes place~\citep{G8.7StarFormation}. The
white contours in Figure~\ref{fig::Hotspot2}~(left) show soft X-ray
emission as detected by the ROSAT satellite~\citep{RosatG8.7}, the
black contours show the 20~cm radio emission as recorded with the
VLA. Another plausible association of HESS\,J1804--216 is the young
Vela-like pulsar PSR\,J1803--2137 with a spin-down age of 16000
years~\citep{PSRJ1803}. The required efficiency of this pulsar's
spin-down luminosity to power the observed emission in \grs\, is only
3\% and it could therefore easily account for the new source.
\subsubsection{HESS\,J1813--178}
\begin{figure}
  \plotone{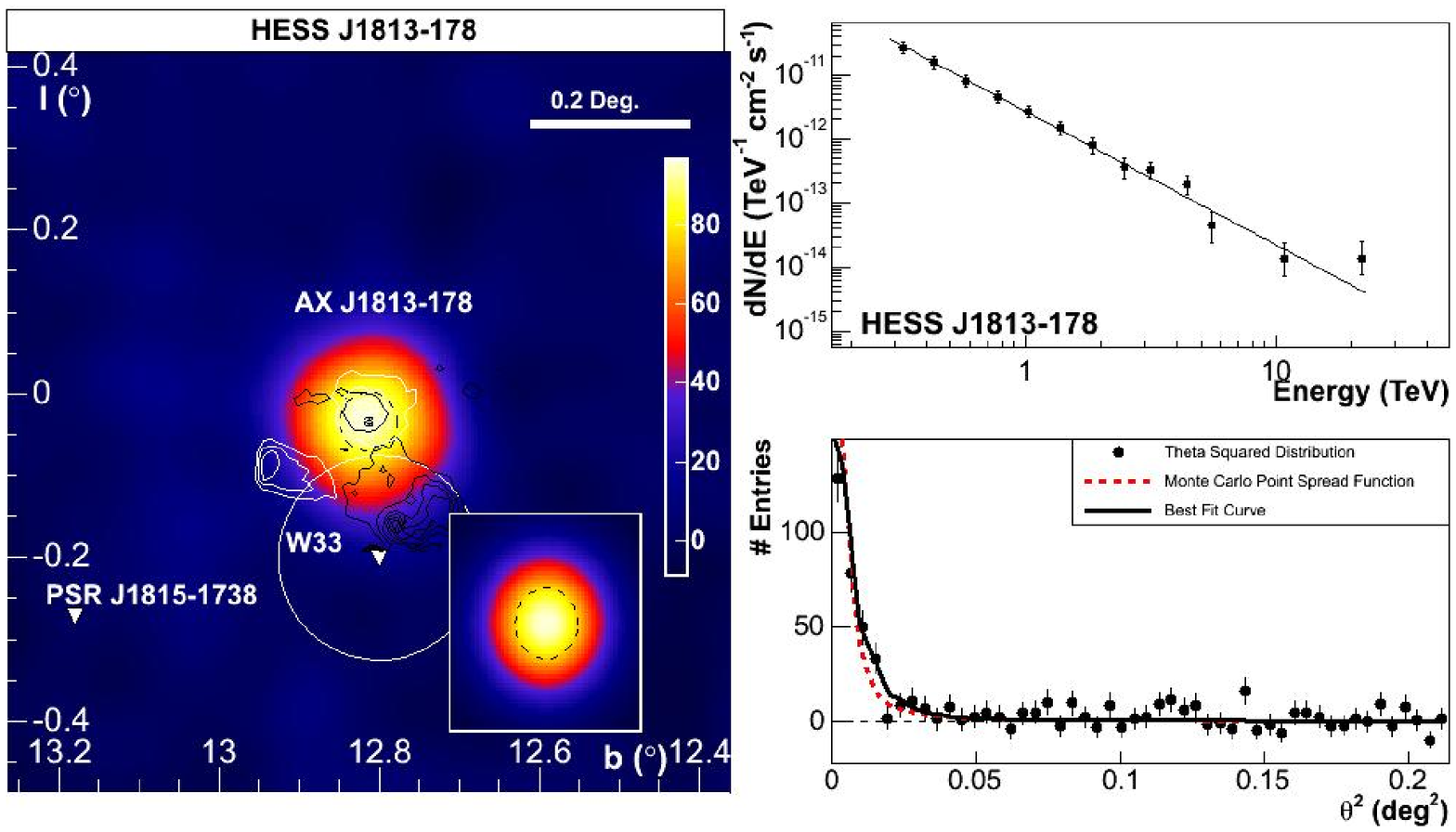}
  \caption{Smoothed excess map (left, smoothing radius 0.04\dg) of the
    region surrounding HESS\,J1813--178 along with nearby pulsars and
    SNRs which were considered as counterparts as discussed in the
    text. The white contours shows the smoothed X-ray count map for
    the region surrounding AX\,J1813--178 obtained by ASCA satellite
    with the GIS instrument in the 0.7--10~keV band, the black
    contours show 20~cm radio emission as observed by the VLA. The
    strong radio source to negative galactic longitudes from
    HESS\,J1813--178 is connected to the W\,33 complex and extends
    further than shown here. The excess seen to negative Galactic
    longitudes at the edge of the ASCA GIS field of view is likely due
    to contamination from GX\,13+1. The radial (lower right)
    distribution ($\theta^2$) plot shows that the source is marginally
    extended with respect to the PSF of the \HESS\ system.
  \label{fig::Hotspot1}}          
\end{figure}

With a moderate flux of approximately 6\% of the Crab Nebula above
200~GeV, this source (see Figure~\ref{fig::Hotspot1}) is among the
more compact of the new sources with an extension of 2.2 arc
minutes. When using a large $\theta^2$ cut of ($0.4^{\circ})^2$, the
normalisation of the power-law fit is 25\% higher, suggesting a faint
non-Gaussian tail in the radial brightness profile. The source
exhibits a rather hard photon index of 2.09$\pm$0.08 (see
Figure~\ref{fig::Hotspot1}~(top right)). In the previous
publication~\citep{HESSScan} this source was still considered as
unidentified. As it is clear from Figure~\ref{fig::Hotspot1}~(left),
there is a compelling positional coincidence between this source and
the X-ray source AX\,J1813--178~\citep{1813Brogan} (white contours in
figure~\ref{fig::Hotspot1} for the 0.7--10~keV band). This ASCA source
is one of the bright sources detected in the ASCA Galactic Plane
survey with a very hard energy spectrum. This object was also recently
detected in the INTEGRAL Galactic plane survey in the 20--100 keV
energy band~\citep{1813Integral}. The region surrounding
HESS\,J1813--178 was also observed with the VLA and 20~cm radio
contours are shown in black in Figure~\ref{fig::Hotspot1}. There is a
faint radio source (G\,12.82--0.02) with a shell-like structure
visible at the position of HESS\,J1813--178 as also recently
reported~\citep{1813Brogan, 1813Helfand}. This indicates a probable
association with a shell-type SNR. Additionally HESS\,J1813--178 lies
10~arc minutes from the centre of the radio source W\,33. W\,33
extends over 15~arc minutes and has a compact radio core
(G\,12.8--0.2) which is 1~arc minute across~\citep{W33Radio}, located
$\sim15$~arc minutes away from HESS\,J1813--178. The W\,33 region
shows signature of an ultra-compact HII region~\citep{UltracompactHII}
and contains methanol, hydroxyl and water masers and other tracers of
recent star formation, that could act as target material for the
generation of \vhe\, \grs\ but is not coincident with the
H.E.S.S. source.
\subsubsection{HESS\,J1825--137}
\begin{figure}
  \plotone{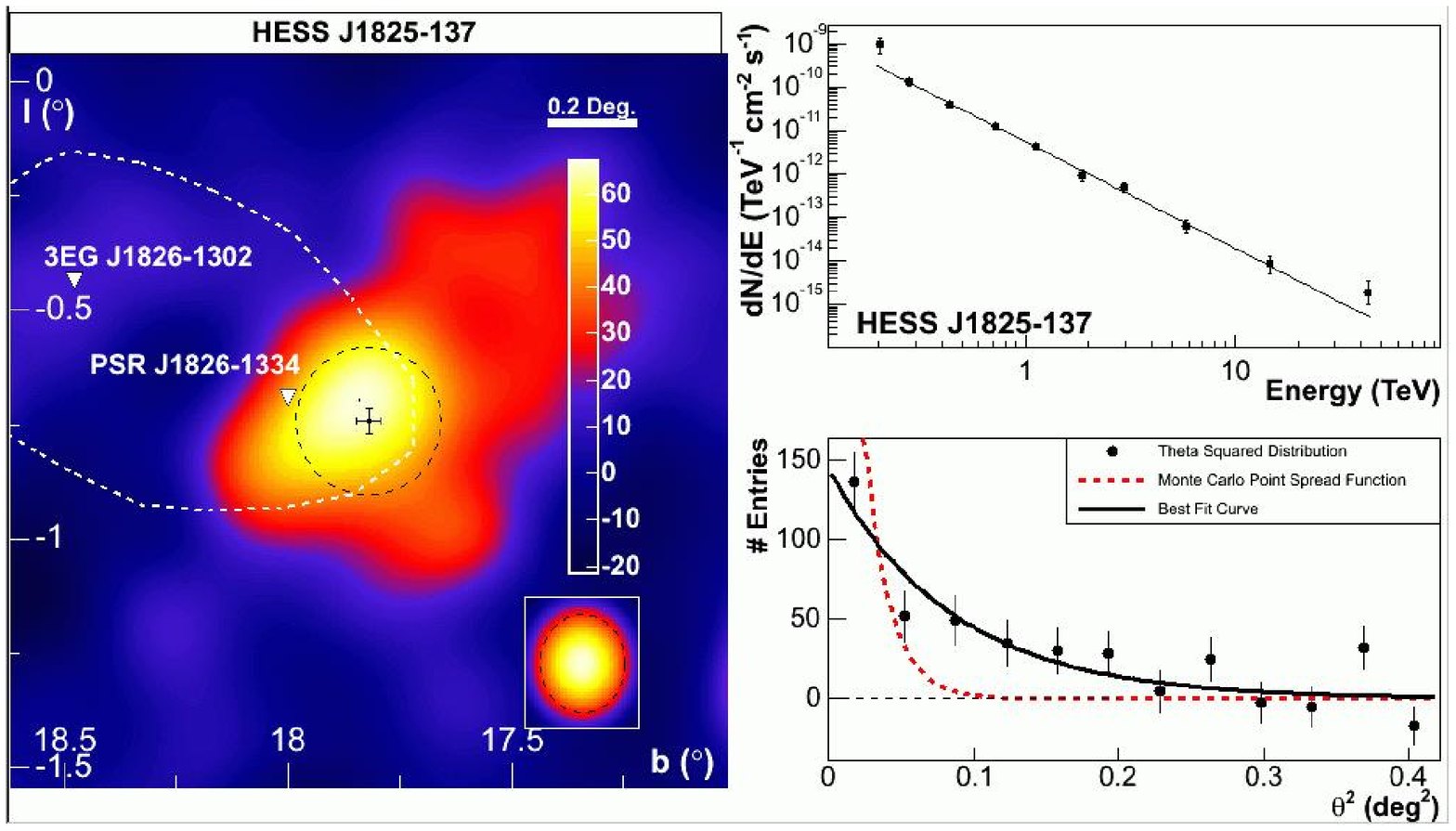}
  \caption{Smoothed excess map (left, smoothing radius 0.09\dg) of the
    region surrounding HESS\,J1825--137 along with the nearby pulsar
    which was considered as counterpart as discussed in the text. The
    differential energy spectrum (top) and the radial distribution
    ($\theta^2$) plot (bottom) are shown on the right hand side.
  \label{fig::Hotspot6}}          
\end{figure}

This source has an energy spectrum with a photon index of
2.46$\pm$0.08 and a flux of \vhe\, \grs\, of approximately 17\% of the
flux from the Crab Nebula above 200~GeV (see
Figure~\ref{fig::Hotspot6}~(top right)). It is approximately radially
symmetric and the emission region can be best described by a Gaussian
of width 10 arc minutes. The position of HESS\,J1825--137 is
approximately 11~arc minutes to negative Galactic longitudes of the
young pulsar PSR\,B1823--13 (PSR\, J1826--1334), which makes an
association of the \vhe\, \gr\, source with the pulsar possible. The
pulsar PSR\,B1823--13 is a 101~ms evolved Vela-like pulsar at a
distance of 3.9$\pm$0.4~kpc and at this distance the VHE luminosity of
HESS\,J1825--137 would only be $\sim$2\% of its spin-down power. The
region surrounding the pulsar has been previously observed at large
zenith angles by the Whipple collaboration~\citep{WhipplePSR1823}, who
observed a 3.1$\sigma$ excess also to negative Galactic longitudes
with respect to the pulsar position. X-ray observations with the
\emph{XMM-Newton} satellite show a 5~arc minutes diameter diffuse
emission region extending also asymmetrically to negative Galactic
longitudes from the pulsar~\citep{XMMPSR1823}. If the \vhe\, \gr\,
source is indeed associated with this pulsar wind nebula, the \vhe\,
emission region would extend much further to the south than in
X-rays. PSR\,B1823--13 was also proposed to power the close-by
unidentified EGRET source 3EG\,J1826--1302~\citep{EGRETPSR1823}. The
distance of the centroid of this EGRET source to HESS\,J1825--137 is
43~arc minutes. As illustrated in Figure~\ref{fig::Hotspot6}~(left),
the new \vhe\, \gr\, source is well compatible with the EGRET source
within the 95\% positional error of EGRET and a simple power-law
extrapolation of the \HESS\ energy spectrum to energies above 1~GeV
gives a similar flux as derived from the EGRET spectrum above this
energy. An in-depth study of HESS\,J1825--137 and dicussion about the
corresponding PWN model will be given in a separate
paper~\citep{HESS1825}.
\subsubsection{HESS\,J1834--087}

\begin{figure}
  \plotone{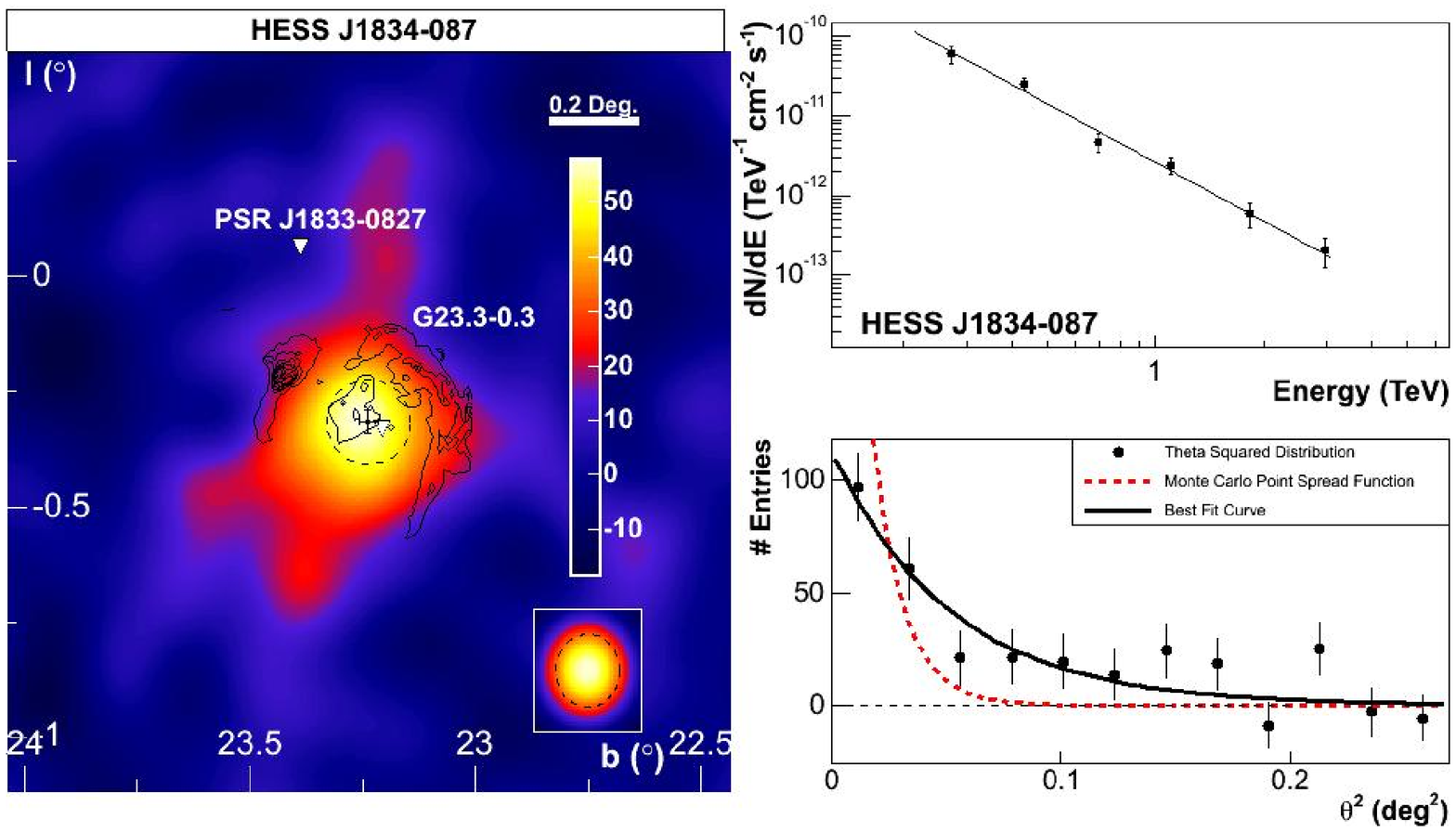}
  \caption{Smoothed excess map (left, smoothing radius 0.08\dg) of the
    region surrounding HESS\,J1834--087 along with nearby pulsars and
    SNRs which were considered as counterparts as discussed in the
    text. The positional coincidence with SNR G23.3--0.3 (W\,41) is
    evident, the black contours show 20~cm radio emission recorded by
    the VLA~\citep{RadioGalacticPlane}. The differential energy
    spectrum (top) and the radial distribution ($\theta^2$) plot
    (bottom) are shown on the right hand side.
  \label{fig::Hotspot5}}          
\end{figure}

HESS\,J1834--087 (Figure~\ref{fig::Hotspot5}) has a size of 12~arc
minutes and a \gr\, flux of 8\% of the flux from the Crab Nebula above
200~GeV. Its energy spectrum exhibits a photon index of 2.45$\pm$0.16
(see Figure~\ref{fig::Hotspot5}~(top right)). When using a large
$\theta^2$ cut of ($0.4^{\circ}$)$^2$, the normalisation of the
power-law fit increases by 30\%, suggesting a deviation of the
morphology from the assumed radial Gaussian brightness profile. The
location of HESS\,J1834--087 makes it another source with a compelling
positional agreement with an SNR (see
Figure~\ref{fig::Hotspot5}~(left)). The positionally coincident
shell-type SNR G23.3--0.3 (W\,41) with a diameter of 27~arc minutes
was mapped in radio with the VLA at 330~MHz~\citep{W41Radio} and at
20~cm~\citep{RadioGalacticPlane} as shown in black contours in
Figure~\ref{fig::Hotspot5}~(left). It is possibly connected to the old
pulsar PSR\,J1833--0827~\citep{W41ConnectionPulsar}. This pulsar is
energetic enough to power HESS\,J1834--087, but its distance of 24~arc
minutes renders an association unlikely; there is also no extended PWN
detected so far.
\subsubsection{HESS\,J1837--069}
\begin{figure}
  \plotone{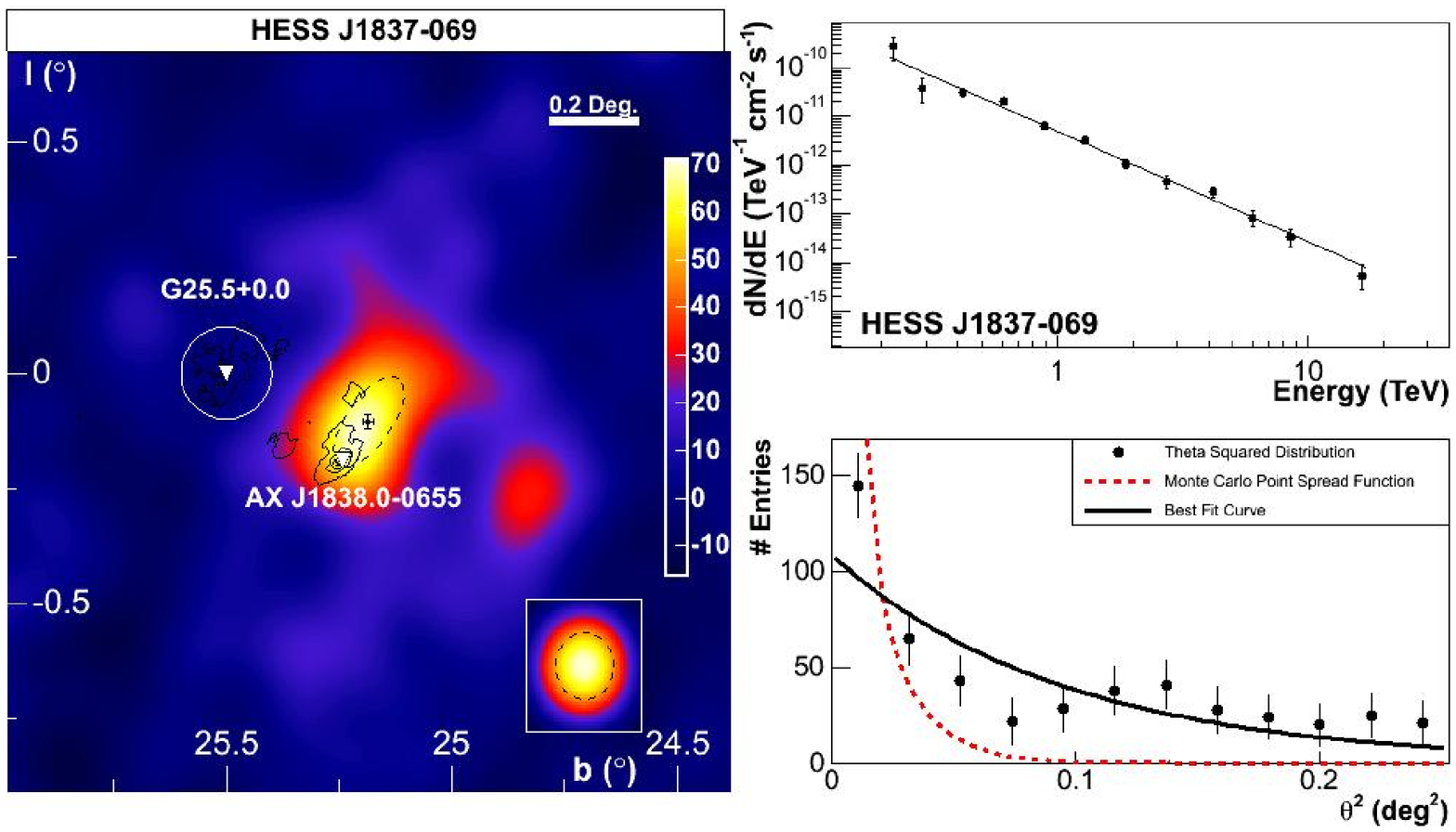}
  \caption{Smoothed excess map (left, smoothing radius 0.05\dg) of the
    region surrounding HESS\,J1837--069 along with nearby SNRs which
    were considered as counterparts as discussed in the text. The
    black contours shows the smoothed X-ray count map for the region
    surrounding AX\,J1838.0--0655 obtained by the ASCA satellite with
    the GIS instrument in the 0.7--10~keV band. The positional
    coincidence of HESS\,J1836--069 with the ASCA source
    AX\,J1838.0--0655 is evident. The differential energy spectrum
    (top) and the radial distribution ($\theta^2$) plot (bottom) are
    shown on the right hand side.
  \label{fig::Hotspot7}}          
\end{figure}

This source (Figure~\ref{fig::Hotspot7}) has an elongated shape with a
semi-major axis of 7$\pm$1~arc minutes and a semi-minor axis of
3$\pm$1~arc minutes. With a flux of 13.4\% of the Crab flux above
200~GeV, the energy spectrum can be well fitted by a power-law of
photon index 2.27$\pm$0.06 (see Figure~\ref{fig::Hotspot7}~(top
right)). When using a large $\theta^2$ cut of ($0.4^{\circ})^2$, the
normalisation of the power-law fit increases by 30\%, suggesting that
the assumption of a Gaussian brightness profile does not hold for this
source. HESS\,J1837--069 coincides with the southern part of the
diffuse hard X-ray complex G25.5+0.0. This source, which is 12~arc
minutes across, was detected in the ASCA Galactic Plane
survey~\citep{BAMBA}. The nature of this bright X-ray source is
unclear but it may be an X-ray synchrotron emission dominated SNR such
as SN\,1006 or a PWN.  The brightest feature in the ASCA map
(AX\,J1838.0--0655), located south of G25.5+0.0, coincides with the
\vhe\, emission. This still unidentified source was also
serendipitously detected by BeppoSAX and also in the hard X-ray
(20--100~keV) band in the Galactic Plane survey of
INTEGRAL~\citep{1838Integral}.
\section{Discussion}
\label{discussion}
Possible multi-wavelength associations of the fourteen new \vhe\,
\gr\, sources were discussed in the previous section and are
summarised in Table~\ref{table::Counterparts}. A further qualification
from tentative associations to unambiguous counterpart identifications
requires spatial or morphological coincidence, a viable \gr\, emission
mechanism for the object, and consistent multi-wavelength behaviour
according to the properties of the suggested identification. Most of
the possible associations given above can be considered as plausible
\gr\, emitters, but only in a minority of these cases does a
satisfactory positional coincidence exist. Four of the new sources
apparently line up with SNRs. These SNRs have been detected and
studied mostly in the radio domain, but in the cases of HESS J1640-465
and HESS J1813-178 X-ray emission was detected by ASCA, and there is
marginal evidence for X-ray emission in ASCA archival data in the case
of HESS J1713-381.  Only in the case of HESS J1834-087 is there
currently no evidence for associated X-ray emission. 

For three of the new sources, namely HESS\,J1616--509,
HESS\,J1804--216 and HESS\,J1825--137, a sufficiently energetic nearby
pulsar could account for the \vhe\, \gr\, emission but the offset of
the sources from the pulsar positions makes these associations less
firm. Three of the new \vhe\, \gr\, sources are consistent with the
95\% positional error of unidentified EGRET sources: HESS\,J1640--465,
HESS\,J1745--303 and HESS\,J1825--137. Two of the new sources are
located very close to a new class of highly absorbed hard X-ray
sources detected recently by INTEGRAL (HESS\,J1632--478 and
HESS\,J1634--472). A third source (HESS\,J1837--067) may be related to
X-ray emission of unknown origin detected by ASCA and INTEGRAL, which
may belong to the same class of objects. The three remaining sources
(HESS\,J1614--518, HESS\,J1702--420 and HESS\,J1708--410) have no
plausible SNR, pulsar or EGRET counterpart and may belong to a new
source class. Significant fluxes of \vhe\, \grs\, without accompanying
X-ray and radio emission would suggest the absence of relativistic
electrons and the presence of energetic nucleons. That such a
population exists is already being suggested by the detection of
apparently ``dark'' accelerators by the HEGRA~\citep{HEGRAUnid} and
\HESS\ ~\citep{HESS1303} collaborations.  However, the possibility
remains that these new sources are associated with SNRs or PWN as yet
undiscovered at other wavelengths.  Further multi-wavelength
investigations are needed to clarify this picture.

The nearest energy band relative to the \HESS\ energy range was
covered by observations with EGRET aboard the Compton Gamma-Ray
Observatory. Its energy coverage was primarily between 30~MeV and
10~GeV and \gr\, source detections above 100~MeV resulted in the third
EGRET (3EG) catalogue~\citep{EGRET}.  The only Galactic sources
detected above 10~GeV are pulsars, for which the sparse photons at
these energies correlate with the location and the pulse profile at
lower EGRET energies~\citep{EGRETPoint}. The bulk of photons above
10~GeV have to be attributed to Galactic (and extra-galactic) diffuse
emission.

When comparing the EGRET sources with the \HESS\ Galactic Plane survey
result, it should be noted that even if an EGRET source spectrum is
measured up to 10~GeV, more than a decade in energy is left uncovered
below the \HESS\ energy threshold. Indeed, only a minor fraction of
the \HESS\ sources coincide within the considerably larger location
uncertainty contours of EGRET sources. These sources are
HESS\,J1640--465, HESS\,J1745--303, and HESS\,J1825--137, which are
consistent with the 95\% location uncertainty of the unidentified
EGRET sources 3EG\,J1639--4702, 3EG\,J1744--3011, and
3EG\,J1826--1302. In total 17 EGRET sources are catalogued within the
\HESS\ survey region. Taking into account the 95\% positional error
contours, they cover an area of 9.76 deg$^2$, resulting in a chance
positional coincidence of a given H.E.S.S. source with an EGRET source
of 32\%. A correlated Galactic latitude distribution of the EGRET and
H.E.S.S. sources, concentrated near the Galactic plane, could increase
this probability.

Physical arguments suggest that the EGRET source population and the
set of new \HESS\ sources might be of different kinds. While the
brightest \gr\, pulsars can be directly traced up to $\sim20$~GeV,
EGRET data indicate the existence of GeV-cutoffs already in the
2--10~GeV range~\citep{EGRETCutoff}. The brightest pulsars, and
several unidentified EGRET sources with pulsar-like characteristics
deviate from single power-law spectral
representations~\citep{EGRETSpectra}, and are significantly better
fitted when introducing high-energy spectral cutoffs.  GeV-cutoffs are
indeed predicted from multi-frequency modelling of the \gr\, emission
of pulsars~\citep{OuterGapPSRs, PolarCapPSRs}. Therefore, the emission
of these unidentified EGRET sources may not extend up to the \HESS\
energy range. From Crab measurements by CELESTE~\citep{CelesteCrab},
consecutively continuing the EGRET GeV-data, a characteristic change
in the emission properties of the pulsar became apparent: pulsed
emission is diminished or entirely absent, leaving only a steady
emission component, in case of Crab attributed to the PWN. If the
majority of EGRET sources in the Galactic Plane may be identifiable
with \gr\, pulsars, this population will overlap with the \HESS\
energies only if a plerionic emission components exist.

For PWN there are theoretical grounds to expect in some cases a
displacement of the \gr\, emission from the pulsar powering the
nebula~\citep{AharonianPWN, Blondin}. In such cases the association
becomes more difficult to establish. Considering only pulsars with
sufficient spin-down luminosity to explain the measured \vhe\, \gr\,
flux, four of the new sources have possible associations with pulsar
wind nebulae. However, only in the case of HESS\,J1825--137, is there
so far a convincing multi-wavelength argument for the offset from the
pulsar position in the sense that an X-ray PWN offset in the source
direction has been detected.

The sensitivity shown in Figure~\ref{fig::Sensitivity} is that for
point-like ($<0.1$\dg) sources. The sensitivity of \HESS\ decreases
with increasing source size:
$F_{\mathrm{min}}\,\sim\,F_{\mathrm{point-source}}\,\sqrt{\sigma_{\mathrm{source}}^{2}
+ \sigma_{\mathrm{PSF}}^{2}}/\sigma_{\mathrm{PSF}}$, where
$\sigma_{\mathrm{PSF}} \sim0.08$\dg\ for the analysis described
here. As a consequence there is a bias in this survey against very
extended sources. In particular, for sources larger than the size of
the ring used for background estimation (0.6\dg\ radius) the flux
threshold increases dramatically since the off-source region used for
the background estimation then also contains flux from the
source. This effect also introduces a bias against close-by sources,
potentially resulting in a latitude distribution narrower than that of
the parent population. An additional consequence of the ring
background technique is that this survey is insensitive to weak
diffuse emission. A search for diffuse emission in this dataset will
be reported elsewhere. Another effect that has an impact on the width
of the latitude distribution is source confusion. Assuming that all
sources lie on the Galactic plane and using the mean RMS size of
$0.1$\dg\ of the new sources, the chance probability of an overlap
between any two of the 17 sources in the scan region is estimated to
be 6\%. Therefore, whilst source confusion could in principle widen
the observed latitude distribution, we expect this effect to be small.

It is apparent from Table~\ref{table::Counterparts} that the most
likely associations of the \gr\, sources lie at rather large distances
(i.e.\ 4--10~kpc) within our Galaxy, and exhibit luminosities in the
range 3--20 $\times10^{34}$~erg~s$^{-1}$. 
Three of the new sources are located in a region of the Galactic plane
that lines up with the 3~kpc arm tangent region (HESS\,J1640--465,
HESS\,J1632--478 and HESS\,J1634--472), which hints at a distance of
8~kpc. HESS\,J1614--518 and HESS\,J1616--508 are at a position in the
plane that lines up with the Norma spiral arm. This arm appears to be
one of the most important massive star forming regions of the
Galaxy~\citep{Bronfman}.

\begin{figure}
  \epsscale{1.0}
  \plotone{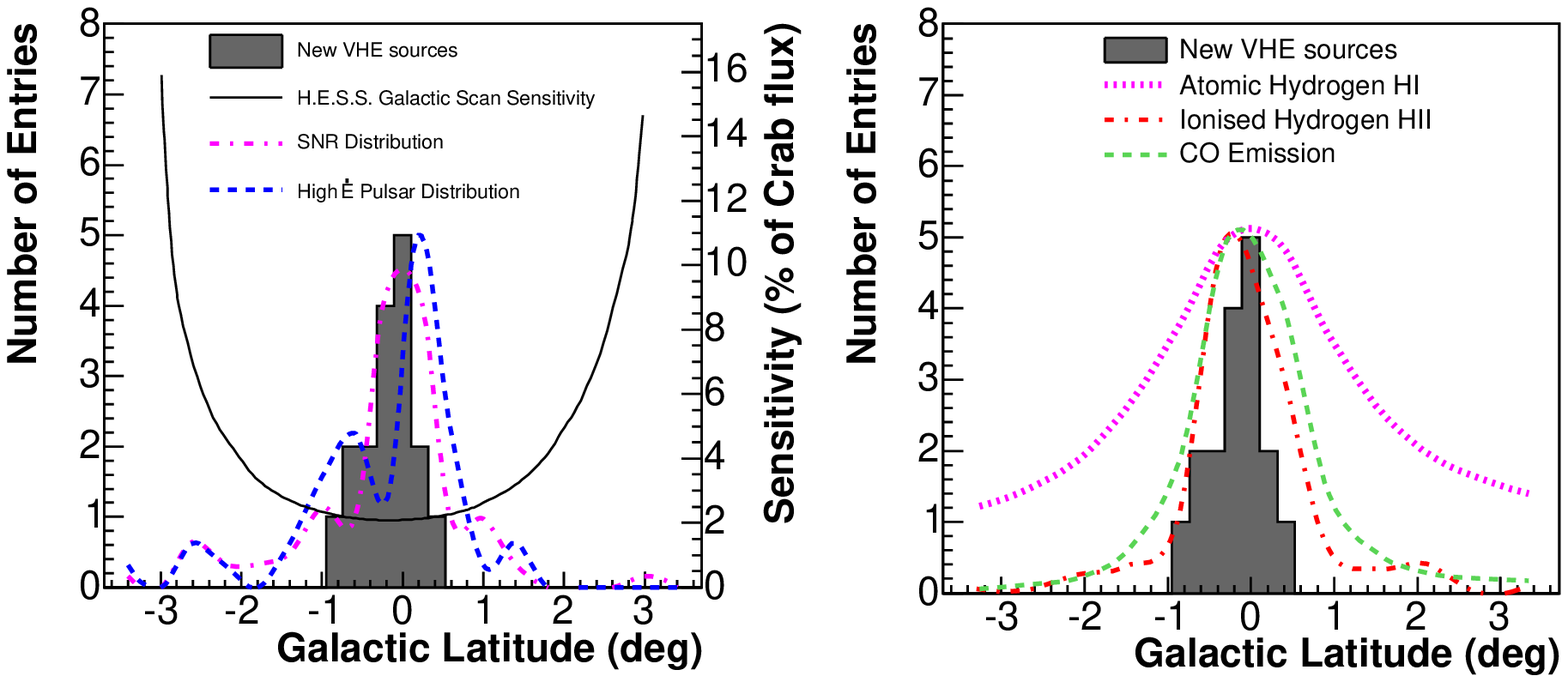}
  \caption{Distribution of Galactic latitude of the seventeen \vhe\,
  \gr\, sources as given in Table~\ref{table::Counterparts} (the
  fourteen new plus the three previously known). The mean of the
  distribution is --0.17 with an RMS of 0.31. On the left, the
  distribution is shown along with the distribution in Galactic
  latitude of possible counterpart populations to the \vhe\, \gr\,
  sources. SNRs from~\citet{Green} are shown in dash-dotted purple,
  high spin-down luminosity pulsars ($\dot E > 10^{34}$~erg~s$^{-1}$)
  taken from the ATNF catalogue~\citep{ATNF} are shown as a dashed
  blue curve. On the right the distribution of \vhe\, \gr\, sources as
  on the left is shown along with the latitude distributions of matter
  in our Galaxy. The distribution of molecular hydrogen
  (HII)~\citep{HIIGalaxy} is shown in dash-dotted red, the
  distribution of atomic hydrogen (HI)~\citep{HIGalaxy} is shown in
  dotted purple and the distribution of CO~\citep{COGalaxy} is shown
  as dashed green line.
  \label{fig::LatDistribution}}
\end{figure}

As it is clear from Table~\ref{table::Counterparts} it is difficult to
explain the new \vhe\, sources as belonging to a single source
population. It is therefore not straight-forward to interpret the
global properties of the sources. However, conventional Galactic
particle accelerators such as young pulsars and SNRs have similar
spatial distribution within the Galaxy. Molecular material and hence
regions of star formation cluster rather close to the Galactic Plane,
with an exponential scale height of $<$100~pc~\citep{Dame}. \gr\,
sources associated with stellar death (and birth) (SNRs, young
pulsars, OB associations) should therefore follow similar
distributions (see Figure~\ref{fig::LatDistribution}). To compare such
candidate populations with the observed Galactic latitude distribution
of the survey sources assumptions must be made about the luminosity
distribution, the radial distribution in the Galaxy, and the intrinsic
size of the sources. A simple Monte-Carlo simulation shows that the
measured $b$ distribution of sources is rather insensitive to the
assumed luminosity distribution.

\begin{figure}
\plotone{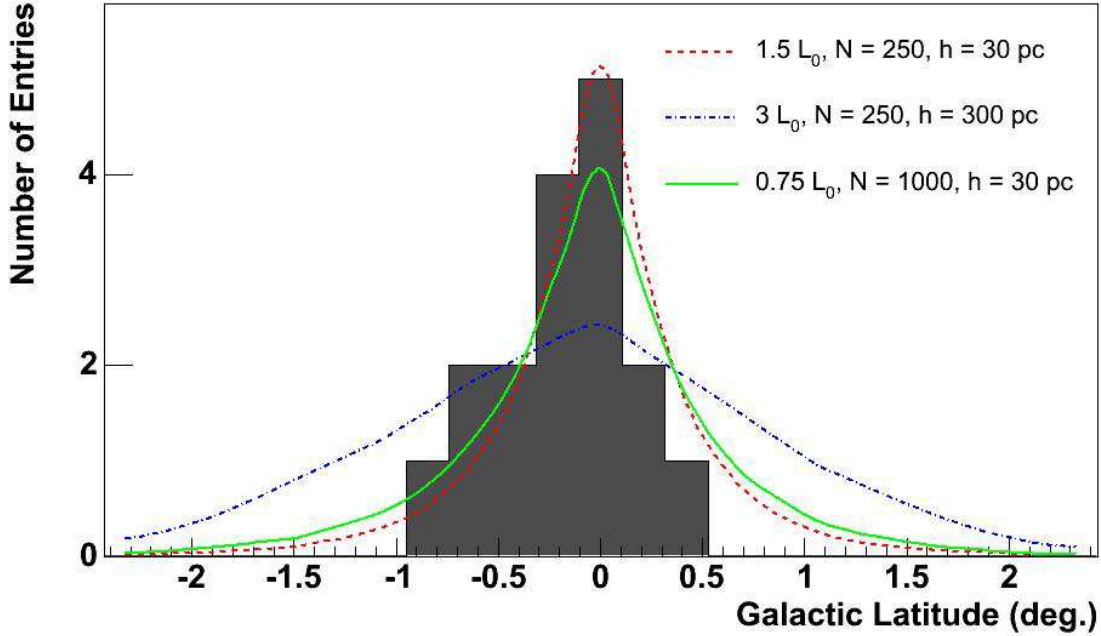}
\caption{Comparison of the Galactic latitude distribution of the
  \vhe\, \gr\, sources in the survey region with the prediction of a
  simple source population model assuming scale heights in the
  Galactic disc of 30~pc (red and green curves) and 300~pc (blue
  curve). The sources are assumed to be mono-luminous. The curves are
  not normalised and give the predicted number of detected sources in
  each latitude bin taking into account the reduced sensitivity at
  higher latitudes~\ref{fig::Sensitivity}. The legend gives the
  physical parameters in each case with $L_0 = 1.3 \times
  10^{34}$~erg~s$^{-1}$ (in the energy range 0.2~TeV to 10~TeV) as
  previously defined. $N$ denotes the number of expected $\gamma$-ray
  sources.}
\label{fig::LatToy}
\end{figure}

Taking the example case of SNRs as being a single class of
counterparts to the new sources, it is possible to derive the expected
properties of the detected \gr\, sources. For this exercise, the
prescription of \citet{DrurySNR} as given in
equation~\ref{equation::SNR} is followed for the \gr\, emission of
SNRs assuming supernova explosions with fixed energy
$10^{51}$~erg~s$^{-1}$ occurring at a rate of 1/40~year$^{-1}$ and
with a \gr\, emitting lifetime of $10^{4}$ years. In this case, 250
\gr\, sources are expected within the Galactic disk. Assuming the
radial distribution of SNRs suggested by \citet{CaseSNR}, the measured
number of the \vhe\, \gr\, sources is consistent with expectations for
plausible values of conversion efficiency into accelerated particles
$\Theta=0.2$ and ISM density $n = 1$~cm$^{-3}$.  As the sensitivity of
the survey is reduced for extended sources, some assumption of the
intrinsic size distribution of sources must be made. Here, the Sedov
solution is assumed to determine the size of the \gr\, emission region
for SNR, of randomly sampled age, expanding into an ISM of density
$n$. Reasonable agreement with the number and angular size of the
detected sources are found for $1 < n <
10$~cm$^{-3}$. Figure~\ref{fig::LatToy} shows the expected
distributions from this simple model for a scale height in the
Galactic disk of 30~pc (red curve) and 300~pc (blue curve). The green
curve is as the red curve but with 4 times as many sources, each with
half the luminosity, illustrating the relative insensitivity of the
shape of the distribution on the assumed source luminosity
distribution.  The difference between the blue and the red curve
illustrates that the \HESS\ survey has a broad enough sensitive
coverage in $b$ to distinguish these two scenarios. For the 300~pc
curve (blue), an intrinsic luminosity twice that of the 30~pc case has
been assumed to give reasonable agreement with the number of detected
sources. As can be seen from Figure~\ref{fig::LatToy} the experimental
data favour a scale height for the parent population of $< 100$~pc in
line with expectations for young SNRs and PWN or massive star forming regions.

\begin{table} 
\begin{center}
\caption{Summary of possible counterparts to the \HESS\ \vhe\, \gr\,
sources in the inner Galaxy. The offset denotes the angular distance
of the \vhe\, \gr\, source from the possible counterpart. Distances
are taken from~\citet{Green, ATNF} and references therein. The last
column gives the implied luminosity between 0.2 and 10~TeV. The three
sources that were known before the \HESS\ survey in the inner Galaxy
(HESS\,J1747--218, HESS\,J1745--290 and HESS\,J1713--397 have been
added for completeness to this table. The possible source classes have
been abbreviated according to the following scheme: Supernova remnant
(SNR), unidentified sources (UID), pulsar wind nebula (PWN), black
hole candidate (BH), X-ray binaries (XRB).}
\vspace{0.5cm}
\label{table::Counterparts} 
\begin{tabular}{|l|c|c|c|c|c|}
\tableline\tableline
Name & Possible & Class & Offset & Distance & Luminosity\\
& Counterpart & & (arcmin) & (kpc) & ($10^{34}$~erg~s$^{-1}$)\\
\tableline
J1614--518 & - & - & - & - & - \\
J1616--508 & PSR\,J1617--5055 & PWN & 10.4 & 6.5 & 20.2\\
J1632--478 & IGR\,J16320--4751 & XRB  & 3  & - & - \\
J1634--472 & IGR\,J16358--4726 / G337.2+0.1& XRB/SNR  & 13/10  & -/14 & -/28.3 \\
J1640--465 & G\,338.3--0.0/3EG\,J1639--4702 & SNR/UID & 0/34 & 8.6 & 16.2 \\
J1702--420 & - & - & - & - & - \\
J1708--410 & - & - & - & - & - \\
J1713--381 & G\,348.7+0.3 & SNR & 0 & 10.2 & 5.2 \\
J1713--397 & RX\,J1713.7--3946 (G\,347.3-0.5) & SNR & 0 & 1 & 1.2 \\
J1745--290 & Sgr A East/Sgr A$^{\star}$ & SNR/BH & 0 & 8.5 & 12.8\\
J1745--303 & 3EG\,J1744--3011 & UID & 10 & - & - \\
J1747--281 & G0.9+0.1 & PWN & 0 & 8.5& 4.4\\
J1804--216 & G\,8.7--0.1/PSR J1803--2137 & SNR/PWN & 21/10.8 & 6/3.9 & 16.5/7.0 \\ 
J1813--178 & G\,12.82--0.02 & SNR & 0 & 4 & 3.1 \\
J1825--137 & PSR\,J1826--1334/3EG\,J1826--1302 & PWN/UID & 11/43 & 3.9/- & 6.1/- \\
J1834--087 & G\,23.3--0.3 & SNR & 0 & 4.8 & 4.4\\
J1837--069 & AX\,J1838.0--0655 & UID & 6 & - & - \\
\tableline 
\end{tabular}
\end{center} 
\end{table}
\section{Conclusions}
\label{conclusion}
The inner part of our Galaxy now contains seventeen known sources of
\vhe\,\grs, compared to three before the \HESS\  survey. The
increased number of sources allows us to consider the behaviour of
\emph{populations} of sources for the first time in this wave-band.  A
major task of multi-wavelength follow-up observations and archival
searches now lies ahead to understand the processes at work in these
astrophysical particle accelerators. The paradigm of cosmic-ray
acceleration in SNRs is consistent with our new findings but it seems
clear that the new sources are not drawn from a single population.
Follow-up observations of the new sources with \HESS\  are planned as
well as an extension of this survey into unexplored regions of the
Galactic Plane.

\acknowledgments
The support of the Namibian authorities and of the University of Namibia
in facilitating the construction and operation of H.E.S.S. is gratefully
acknowledged, as is the support by the German Ministry for Education and
Research (BMBF), the Max Planck Society, the French Ministry for Research,
the CNRS-IN2P3 and the Astroparticle Interdisciplinary Programme of the
CNRS, the U.K. Particle Physics and Astronomy Research Council (PPARC),
the IPNP of the Charles University, the South African Department of
Science and Technology and National Research Foundation, and by the
University of Namibia. We appreciate the excellent work of the technical
support staff in Berlin, Durham, Hamburg, Heidelberg, Palaiseau, Paris,
Saclay, and in Namibia in the construction and operation of the
equipment. We thank Y. Uchiyama for assistance with the ASCA
analysis. This research has made use of data obtained from the High
Energy Astrophysics Science Archive Research Center (HEASARC),
provided by NASA's Goddard Space Flight Center.

\end{document}